\documentclass[pre,onecolumn]{revtex4-2}

\usepackage{graphicx}
\usepackage[english]{babel}
\usepackage{amsmath}
\usepackage{amssymb}
\usepackage{siunitx}
\usepackage[colorlinks=true, allcolors=blue]{hyperref}
\usepackage{cleveref}

\renewcommand{\selectlanguage}[1]{} 

\newcommand{\lp}{\left(}
\newcommand{\rp}{\right)}
\newcommand{\lbk}{\left[}
\newcommand{\rbk}{\right]}
\newcommand{\nn}{\nonumber}
\newcommand{\di}{\mathrm{d}}

\begin{document}

\title{From noisy cell size control to population growth: \\ when variability can be beneficial}
\author{Arthur Genthon}
\affiliation{Max Planck Institute for the Physics of Complex Systems, 01187 Dresden, Germany}

\begin{abstract}

Single-cell experiments revealed substantial variability in generation times, growth rates but also in birth and division sizes between genetically identical cells. 
Understanding how these fluctuations determine the fitness of the population, i.e. its growth rate, is necessary in any quantitative theory of evolution.
Here, we develop a biologically-relevant model which accounts for the stochasticity in single-cell growth rates, birth sizes and division sizes. 
We derive expressions for the population growth rate and for the mean birth size in the population in terms of the single-cell fluctuations.
Allowing division sizes to fluctuate reveals how the mechanism of cell size control (timer, sizer, adder) influences population growth. 
Surprisingly, we find that fluctuations in single-cell growth rates can be beneficial for population growth when 
slow-growing cells tend to divide at smaller sizes than fast-growing cells. 
Our framework is not limited to exponentially-growing cells like \textit{Escherichia coli}, and we derive similar expressions for cells with linear and bi-linear growth laws, such as \textit{Mycobacterium tuberculosis} and fission yeast \textit{Schizosaccharomyces pombe}, respectively.

\end{abstract} 

\maketitle

\section{Introduction}

In good conditions with an adequate supply of resources, the number $N$ of cells in a population grows exponentially in the long time limit, $N \sim e^{\lambda t}$, where the population growth rate $\lambda$, or Malthus parameter, is a proxy for population fitness. It is well understood that this fitness depends on the environmental conditions, such as temperature and nutrients for instance \cite{monod_growth_1949,schaechter_dependency_1958,ratkowsky_relationship_1982}. 
However, recent single-cell experiments revealed a high degree of phenotypic variability between individual cells of isogenic populations, for example in generation times (time from cell birth to division) and single-cell growth rates, even in steady environments \cite{wang_robust_2010,osella_concerted_2014,campos_constant_2014,taheri-araghi_cell-size_2015,tanouchi_noisy_2015,hashimoto_noise-driven_2016}. The dependence of the population growth rate on this inter-cellular variability is starting to be investigated \cite{hashimoto_noise-driven_2016,thomas_single-cell_2017,olivier_how_2017,lin_effects_2017,lin_optimal_2019,lin_single-cell_2020,barber_modeling_2021} but is yet to be fully understood.

In a seminal article, Powell showed that variability in generation times increases the population growth rate for age-regulated populations \cite{powell_growth_1956}, a control mechanism today called timer \cite{jun_fundamental_2018}. However, the timer mechanism does not account for the observed mother-daughter correlations in generation times \cite{lin_effects_2017}, and also does not lead to a steady size distribution for exponentially-growing cells \cite{bell_cell_1968,trucco_note_1970}. Steady-state size distributions are reached when cell size is controlled instead of cell age.
Mechanisms of cell size control include the sizer strategy, where cells divide when reaching a critical size, and the adder strategy where cells divide after growing by a critical added size since birth \cite{jun_fundamental_2018}. 
Moreover, these mechanisms produce the expected negative mother-daughter correlations in generation times, since, intuitively, a cell with a large generation time tends to divide big and thus to give birth to a big daughter cell, which in turn takes a short time to reach its target division size \cite{taheri-araghi_cell-size_2015,lin_effects_2017}. 
Experiments showed that many organisms indeed follow these phenomenological size control strategies \cite{jun_fundamental_2018}, with an ubiquity of near-adder mechanisms in bacteria such as \textit{Escherichia coli} \cite{campos_constant_2014,taheri-araghi_cell-size_2015,tanouchi_noisy_2015}, \textit{Caulobacter crescentus} \cite{campos_constant_2014}, \textit{Bacillus subtilis} \cite{taheri-araghi_cell-size_2015}; budding yeast \textit{Saccharomyces cerevisiae} \cite{soifer_single-cell_2016}; \textit{Mycobacterium smegmatis} \cite{priestman_mycobacteria_2017}; archaeal cell \textit{Halobacterium salinarum} \cite{eun_archaeal_2017} and in a variety of mammalian cells \cite{cadart_size_2018}, while a near-sizer strategy characterizes the fission yeast \textit{Schizosaccharomyces pombe} \cite{nobs_long-term_2014}.

Recent re-examination of the impact of single-cell variability on population growth, for size-regulated populations of exponentially-growing cells, showed that population fitness is primarily set by the statistics of single-cell growth rates rather than generation times \cite{thomas_single-cell_2017,lin_effects_2017}. Indeed, in the absence of fluctuations in single-cell growth rates, the population grows at the same rate as its individual cells independently of the presence of other sources of noise.
This is a consequence of cell size control and can be understood intuitively as follows. Cell size grows as $s(t)=s_b e^{\alpha t}$ with $s_b$ the size at birth and $\alpha$ the single-cell growth rate, 
therefore the total population volume $V(t)=\sum_{i=1}^{N(t)} s_i(t)$ grows as $V(t)=V_0 e^{\alpha t}$ when $\dot{s}_i=\alpha s_i$ with the same $\alpha$ for all cells $i$. Since cell size is regulated, the average cell size in a population snapshot, $V/N \sim e^{(\alpha-\lambda)t}$, must converge, which implies $\lambda=\alpha$. 
On the other hand, population fitness is lowered by uncorrelated fluctuations in single-cell growth rates, because slow-growing cells have larger than average generation times and are thus more likely to appear in population snapshots, thus biasing the instantaneous growth rate of the total population volume towards low values \cite{thomas_single-cell_2017,lin_effects_2017}. 

In these works, only the fluctuations in the single-cell growth rates have been investigated, and the mechanism of cell size control was considered infinitely precise such that all cells are born with size $s_b=1$ and divide with size $s_d=2$. Single-cell data nonetheless indicate that cell size control is a noisy process, where cells born with the same size can divide at different sizes, and where the two cells born from a division can be born with different birth sizes \cite{osella_concerted_2014,campos_constant_2014,taheri-araghi_cell-size_2015,tanouchi_noisy_2015,eun_archaeal_2017,priestman_mycobacteria_2017}. The roles of cell size control and of noises in cell size for population growth thus remain to be studied.

In this article, we propose an approach to investigate how population growth depends on fluctuations at the single-cell level, namely in growth rates, in cell size control and in size partitioning upon division. We start by computing the corrections to the population growth rate coming from noise in cell size control and in size partitioning for exponentially-growing cells, thus generalizing the result from \cite{thomas_single-cell_2017,lin_effects_2017}. Those corrections are of second order, as they are conditioned on the presence of noise in single-cell growth rates, and depend explicitly on the mechanism of cell size control. Our approach is not limited to exponentially-growing cells, and we derive similar expressions for other biologically-relevant growth laws such as linear growth, recently observed in \textit{Mycobacterium tuberculosis} \cite{chung_mycobacterium_2023} and the bi-linear growth pattern of fission yeast \textit{S. Pombe} \cite{horvath_cell_2016}. In these cases, corrections due to noise in cell size control and in size partitioning are of first order and independent of the presence of noise in single-cell growth rates. Finally, our analysis of data from \cite{taheri-araghi_cell-size_2015} for \textit{E. coli} in different conditions indicates that the mechanism of cell size control itself depends on the single-cell growth rate for exponentially-growing cells. 
In this case, we 
show that, unlike what was observed in \cite{thomas_single-cell_2017,lin_effects_2017}, uncorrelated fluctuations in single-cell growth rates can be beneficial for population growth.

\section{Model}
\label{sec:model}

\subsection{Population dynamics}

\begin{figure*}
    \centering
    \includegraphics[width=\linewidth]{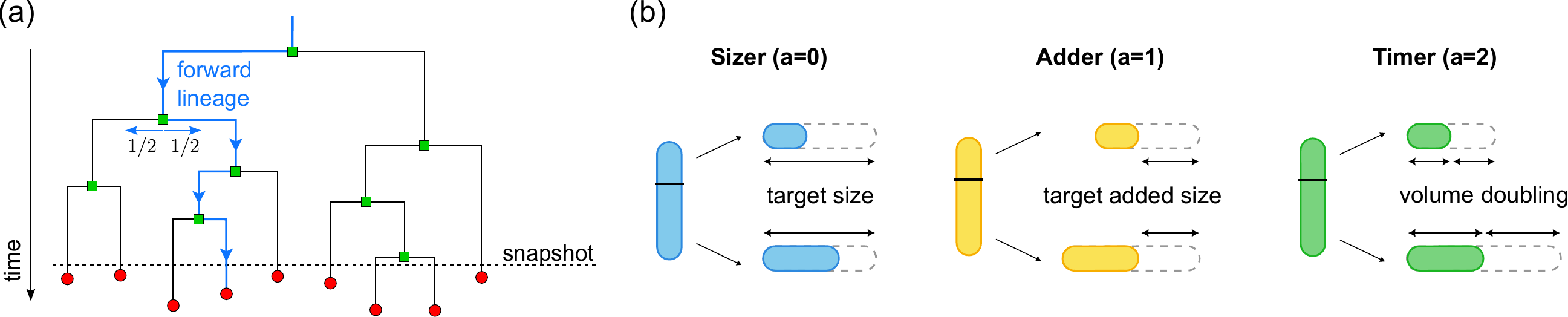}
    \caption{Schematics of the population tree and the mechanisms of cell size control. (a) Population tree starting with one ancestor cell. Green squares and red circles indicate divisions that occurred before and after the population snapshot at time $t$ (dashed line), respectively. Cells present in the population at time $t$ are called leaf cells. A forward lineage obtained when tracking only one of the two daughter cells at each division with equal probability $1/2$ is highlighted in blue. (b) Models of cell size control, where the filled cells and the empty cells with dashed contours represent the size at birth and the target size at division, respectively. Each mechanism is captured by a value of the slope $a$ in the noisy linear map \cref{eq_csc}. For the timer, cells double their volume on average only for single-cell exponential growth.}
    \label{fig:tree_csc}
\end{figure*}

Throughout the article, we use the terms size and volume interchangeably. For rod-shaped organisms like \textit{E. coli} which maintain an approximately constant width during growth, length is a proxy of volume. Moreover, we consider that single-cell growth rates are constant during the cell cycle. The effect of time-varying single-cell growth rates on population growth and on the cell size distribution have been investigated in \cite{levien_non-genetic_2021,genthon_analytical_2022,hein_asymptotic_2024}.

In a freely-growing population, the expected number of cells $n(\tau,s_d,s_b,\alpha;t)$ with age $\tau$, division size $s_d$, birth size $s_b$ and growth rate $\alpha$ in the population at time $t$ follows:
\begin{equation}
\label{eq:pbe_alpha}
	\partial_t n(\tau,s_d,s_b,\alpha;t) = - \partial_{\tau}  n(\tau,s_d,s_b,\alpha;t)
\end{equation}
for $0 < \tau \leq \tau_d$, where $\tau_d$ is the generation time (age at division). For exponentially-growing cells $\tau_d=\ln(s_d/s_b)/\alpha$, and for linearly-growing cells $\tau_d=(s_d-s_b)/\alpha$, for example.
Please note that we treat $s_d$ as a variable for mathematical convenience, and that the same results are obtained when using a division rate instead, see Supplementary Material.

The boundary condition for newborn cells reads:
\begin{equation}
\label{eq:bc_alpha}
    n(\tau=0,s_d,s_b,\alpha;t) = 2 \int \di s_d' \di s_b' \di \alpha' \ 
 \mathcal{K}(s_d,s_b,\alpha|s_d',s_b',\alpha')n(\tau_d',s_d',s_b',\alpha';t) \,,
\end{equation}
where $\mathcal{K}(s_d,s_b,\alpha|s_d',s_b',\alpha')$ is the probability for a newborn cell to inherit a size  at birth $s_b$, a size at division $s_d$ and an single-cell growth rate $\alpha$, knowing its mother was born with size $s_b'$ and single-cell growth rate $\alpha'$ and divided at size $s_d'$. 

In the steady-state limit, the population grows as:
\begin{equation}
\label{eq:n_ss}
n(\tau,s_d,s_b,\alpha;t) \underset{t \to \infty}{\sim} \Pi(\tau,s_d,s_b,\alpha) e^{\lambda t} \,,
\end{equation}
where $\Pi(\tau,s_d,s_b,\alpha)$ is the joint snapshot distribution of age, division size, birth size and single-cell growth rate, and $\lambda$ the population growth rate. The snapshot distribution is obtained by sampling the cells present in the population at a given snapshot time $t$, as shown with the dashed line in \cref{fig:tree_csc}a, and becomes independent of time in the steady-state limit. 
When inserting \cref{eq:n_ss} into \cref{eq:pbe_alpha}, one solves the equation for the joint distribution:
\begin{equation}
\label{eq_Pi_sol}
    \Pi(\tau,s_d,s_b,\alpha) = \begin{cases}
        \Pi(0 ,s_d,s_b,\alpha) e^{-\lambda \tau} & \tau \leq \tau_d \\
        0 & \tau > \tau_d \,.
    \end{cases}
\end{equation}

Finally, we define the joint tree distribution 
\begin{equation}
	\label{eq_def_rho}
    \rho_{\mathrm{tree}}(s_d,s_b,\alpha)= \frac{\Pi(0 ,s_d,s_b,\alpha)}{\int \di s_d' \di s_b' \di \alpha' \Pi(0 ,s_d',s_b',\alpha')} \,,
\end{equation}
which is the distribution for newborn cells throughout the whole tree, including cells that are currently in the population and have not divided yet (leaf cells, see \cref{fig:tree_csc}a). 

Using \cref{eq:n_ss,eq_Pi_sol,eq_def_rho}, we recast the boundary condition for newborn cells as a generalized Euler-Lotka equation:
\begin{equation}
	\label{eq_main_powell}
	\rho_{\mathrm{tree}}(s_d,s_b,\alpha) = 2  \int \di s_d' \di s_b' \di \alpha' \ 
	\mathcal{K}(s_d,s_b,\alpha|s_d',s_b',\alpha') 
	 \rho_{\mathrm{tree}}(s_d',s_b',\alpha') e^{-\lambda \tau_d(s_d',s_b',\alpha')}\,,
\end{equation}
which relates the population growth rate $\lambda$ to single-cell variability. 

Let us first illustrate this result for exponentially-growing cells in the case of perfect size control, when in steady state all cells are born with size $s_b=1$ and divide at size $s_d=2$. The only source of variability therefore comes from fluctuations in single-cell growth rates, and the kernel reduces to $\mathcal{K}(s_d,s_b,\alpha|s_d',s_b',\alpha') = \nu(\alpha)\delta(s_d - 2 s_b) \delta(s_b - s_d'/2)$, with $\delta$ the Dirac delta function and $\nu(\alpha)$ the distribution of single-cell growth rates. We assume no mother-daughter correlations in single-cell growth rates, yet mother-daughter correlations in generation times are present as a result of cell size control, as explained in the introduction.
Integrating \cref{eq_main_powell} over $s_d$, $s_b$ and $\alpha$ gives the simple Euler-Lotka equation $1= \int \di \alpha \ 2^{1-\lambda/\alpha} \nu(\alpha)$, from which an expansion at leading order for small noise in $\alpha$ gives \cite{thomas_single-cell_2017,lin_effects_2017}:
\begin{equation}
\label{eq:lam_2nd_order}
    \frac{\lambda}{\langle \alpha \rangle} \approx  1 - \lp 1 -\frac{\ln 2}{2} \rp \mathrm{CV}_{\alpha}^2 \,,
\end{equation}
with $\mathrm{CV}_{\alpha}$ the Coefficient of Variation (CV=standard deviation/mean) of distribution $\nu(\alpha)$.
In this case, fluctuations in single-cell growth rates are detrimental for population growth. Note that fluctuations can be beneficial when considering mother-daughter correlations in single-cell growth rates with kernel $\nu(\alpha|\alpha')$ \cite{lin_single-cell_2020}. In the following, we do not consider such correlations. 

In general however, size control is not infinitely precise and the population growth rate and the size distribution, which are both unknown, are entangled by \cref{eq_main_powell}. 
In the following sections, we will derive explicit expressions for the population growth rate and the mean birth size by using \cref{eq_main_powell} in the small noise limit. 

To fully characterize the population dynamics, we now need to detail the inheritance laws at division.
In the remaining of the article, we consider the following class of kernels: 
\begin{equation}
\label{eq:K_fact}
    \mathcal{K}(s_d,s_b,\alpha|s_d',s_b',\alpha') = \psi(s_d|s_b,\alpha) \nu(\alpha) \kappa(s_b|s_d') \,.
\end{equation}
The size at division of the newborn cell depends on its size at birth and its growth rate via the distribution $\psi(s_d|s_b,\alpha)$. This distribution defines the strategy of cell size control, as detailed in a later section. The growth rate of the newborn cell is randomly drawn from distribution $\nu(\alpha)$. The variability in volume partitioning at division is captured by the kernel $\kappa(s_b|s_d')$. We further focus on the case where volume partitioning depends only on the daughter-to-mother volume fraction, irrespective of the absolute volumes:
\begin{equation}
	\kappa(s_b|s_d')=\int_0^1 \di p \ \pi(p) \delta(s_b- p s_d') \,,
\end{equation}
where $\pi(p)$ is the probability for a daughter cell to inherit a fraction $p$ of its mother volume at division. Volume conservation imposes $\langle p \rangle = 1/2$, and noise in volume partitioning is measured by $\mathrm{CV}_p=2 \sigma_p$.

The factorized form of kernel $\mathcal{K}$, \cref{eq:K_fact}, implies that $\rho_{\mathrm{tree}}(s_d|s_b,\alpha)=\psi(s_d|s_b,\alpha)$, and that the size at birth and the single-cell growth rate are independent random variables in the tree statistics: $\rho_{\mathrm{tree}}(s_b,\alpha)=\rho_{\mathrm{tree}}(s_b)\rho_{\mathrm{tree}}(\alpha)$ with $\rho_{\mathrm{tree}}(\alpha)=\nu(\alpha)$.

\subsection{Forward lineages}

It will be useful in the Results sections to consider the statistics obtained when following a single lineage, like in mother machine experiments where isolated lineages are monitored \cite{wang_robust_2010}. This statistics can also be obtained in freely-growing populations by sampling the lineages of the population tree with non-uniform weights given by $2^{-D_l}$ where $D_l$ is the number of divisions along lineage $l$ \cite{nozoe_inferring_2017}. This weight is derived when following a lineage forward in time and choosing with probability $1/2$ which of the two daughter cells to track at each division. Therefore, the isolated lineage statistics is also called the forward statistics. An example of forward lineage is highlighted in blue in \cref{fig:tree_csc}a. 

In the forward statistics, only one daughter cell is followed at division (instead of two in the tree statistics) and the number of cells is constant, so that the boundary condition for newborn cells in the forward statistics is obtained by removing the factor $2$ and setting $\lambda=0$ in \cref{eq_main_powell}:
\begin{equation}
\label{eq_BC_fw}
 \rho_{\mathrm{fw}}(s_d,s_b,\alpha) =  \int \di s_d' \di s_b' \di \alpha' \ 
 \mathcal{K}(s_d,s_b,\alpha|s_d',s_b',\alpha') 
 \rho_{\mathrm{fw}}(s_d',s_b',\alpha') \,.
\end{equation}
Like for the tree distribution, the factorized form of kernel $\mathcal{K}$, \cref{eq:K_fact}, implies that $\rho_{\mathrm{fw}}(s_d|s_b,\alpha)=\psi(s_d|s_b,\alpha)$, and that the size at birth and the single-cell growth rate are independent random variables in the forward statistics: $\rho_{\mathrm{fw}}(s_b,\alpha)=\rho_{\mathrm{fw}}(s_b)\rho_{\mathrm{fw}}(\alpha)$ with $\rho_{\mathrm{fw}}(\alpha)=\nu(\alpha)$. 

In the rest of the article, we therefore note $\langle \alpha \rangle$ and $\mathrm{CV}_{\alpha}$ the average value and coefficient of variation of single-cell growth rates with respect to distribution $\nu$, equivalently measured in the tree or forward statistics.

\subsection{Cell size control}

Recent works have shown that the size control of many organisms obeys a phenomenological noisy linear map, which is a discrete description where the cell size at division is related to the cell size at birth by \cite{tanouchi_noisy_2015,jun_cell-size_2015}:
\begin{equation}
	\label{eq_csc}
    s_d = a s_b + b + \eta \,,
\end{equation}
with $0 \leq a \leq 2$, $b \geq 0$ and $\eta$ a noise such that $\langle \eta \rangle=0$.
The linear map captures the sizer mechanism when $a=0$ and $b$ is the target division size, the adder mechanism when $a=1$ and $b=\Delta$ is the target added size between birth and division, and the timer mechanism for single-cell exponential growth when $a=2$ and $b=0$. These mechanisms are illustrated in \cref{fig:tree_csc}b.
Since the timer mechanism is not compatible with single-cell exponential growth \cite{bell_cell_1968,trucco_note_1970}, in the following we assume that $0 \leq a <2$ and $b>0$. The consistency of these definitions of the sizer and adder with those based on division rates is detailed in the Supplementary Material.

The noisy linear map (\ref{eq_csc}) implies that $\psi(s_d|s_b,\alpha)=\varphi(s_d-a s_b|\alpha)$. Based on mother-machine data analyses suggesting that the added size is independent of the single-cell growth rate \cite{taheri-araghi_cell-size_2015}, it is usually assumed in the literature that the parameters $a$, $b$ and $\eta$ are independent of the single-cell growth rate $\alpha$, so that $\varphi(s_d-a s_b|\alpha) = \varphi(s_d-a s_b)$. In this case, we use $\mathrm{CV}_{\varphi}^2=(\langle \varphi^2 \rangle_{\mathrm{fw}} - \langle \varphi \rangle_{\mathrm{fw}}^2)/ \langle \varphi \rangle_{\mathrm{fw}}^2 = \langle \eta^2 \rangle_{\mathrm{fw}} / b^2$ to quantify the noise in cell size control. 
In particular, for the adder $\mathrm{CV}_{\varphi}$ is the CV of the added volume $\Delta_d=s_d-s_b$ between birth and division, $\mathrm{CV}_{\varphi}=\mathrm{CV}_{\mathrm{fw}}[\Delta_d]$, and for the sizer it is equal to the CV of the size at division, $\mathrm{CV}_{\varphi}=\mathrm{CV}_{\mathrm{fw}}[s_d]$.
In the following, we assume that the noisy linear map, and thus the target division size, are independent of the single-cell growth rate, and we relax this assumption later in the Results section.

\section{Results}

\subsection{Noisy cell size control and volume partitioning for exponentially-growing cells}
\label{sec:noisy_control}

\begin{figure*}
    \centering
    \includegraphics[width=\linewidth]{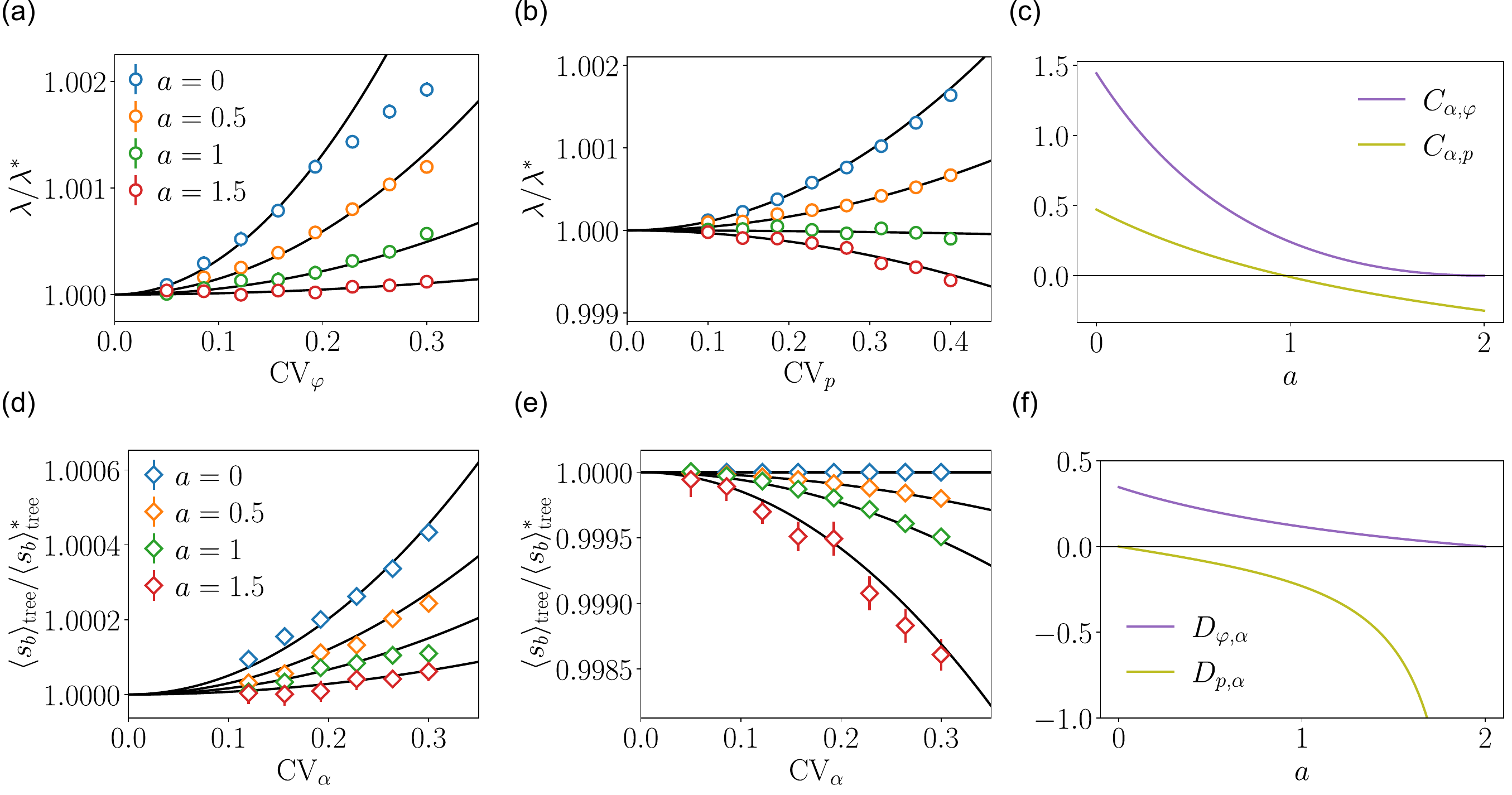}
    \caption{Impact of single-cell variability on population growth and mean birth size for exponentially-growing cells. 
    Noises in size control $ \mathrm{CV}_{\varphi}$, in volume partitioning $\mathrm{CV}_{p}$ and in single-cell growth rate $\mathrm{CV}_{\alpha}$ are varied, for different models of cell size control: sizer ($a=0$), intermediate between sizer and adder ($a=0.5$), adder ($a=1$) and intermediate between adder and timer, or timer-like, ($a=1.5$). 
    $\lambda$ and $\langle s_b \rangle_{\mathrm{tree}}$ are re-scaled by their values $\lambda^*$ and $\langle s_b \rangle_{\mathrm{tree}}^*$ in the absence of fluctuations in the variable on the $x$-axis ($\mathrm{CV}=0$), given explicitly in the Methods section up to order $\mathrm{CV}^4$ for the other variables.
    Theoretical predictions up to order $\mathrm{CV}^4$ are shown in black and dots and diamonds are the results of numerical simulations. The determination of error bars and other details on the simulations are explained in the Methods section.
    (a,b) $\mathrm{CV}_{\alpha}=0.15$, (a) $\mathrm{CV}_{p}=0$, (b) $\mathrm{CV}_{\varphi}=0$, (d) $\mathrm{CV}_{\varphi}=0.12$, $ \mathrm{CV}_{p}=0$, (e) $ \mathrm{CV}_{p}=0.16$, $\mathrm{CV}_{\varphi}=0$, (a,b,d,e) $b=1$, $\langle \alpha \rangle =1$. (c,f) Coefficients in the small noise expansions \cref{eq_lambda_exp_coeffs,eq_sb_exp_coeffs} against the mechanism of cell size control $a$.}
    \label{fig:lambda_vs_noise}
\end{figure*}

Exponential growth is the most common growth law amongst mircro-organisms, reported for many bacterial cells  \cite{wang_robust_2010,robert_division_2014,osella_concerted_2014,campos_constant_2014,taheri-araghi_cell-size_2015}, and also for budding yeast \textit{S. cerevisiae} \cite{soifer_single-cell_2016}, for archaeal cell \textit{H. salinarum} \cite{eun_archaeal_2017}, and for mycobacterium \textit{M. smegmatis} \cite{priestman_mycobacteria_2017}, for example. In this section, we show how population growth depends on noise in size control and volume partitioning for exponentially-growing cells.

For those cells, as explained in the introduction, corrections to the population growth rate due to noise in cell size control and in volume partitioning must enter as terms which couple to noise in single-cell growth rates in the small noise limit \cite{lin_effects_2017}:
\begin{equation}
	\label{eq_lambda_exp_coeffs}
    \frac{\lambda}{\langle \alpha \rangle} \approx 1 + C_{\alpha} \mathrm{CV}_{\alpha}^2 + C_{\alpha, \varphi} \mathrm{CV}_{\alpha}^2 \mathrm{CV}_{\varphi}^2 + C_{\alpha, p} \mathrm{CV}_{\alpha}^2 \mathrm{CV}_{p}^2 \,.
\end{equation}
Here and in the rest of the article, we present expansions up to the leading order term for each source of noise. For example, terms in $\mathrm{CV}_{\alpha}^3$ and $\mathrm{CV}_{\alpha}^4$ are omitted because the correction coming from fluctuating growth rates is controlled by the term in $\mathrm{CV}_{\alpha}^2$. When comparing predictions to numerical simulations on the other hand, we keep corrections up to a consistent order, as detailed later.
To determine the coefficients $C$ we need to Taylor expand \cref{eq_main_powell}, which entangles $\lambda$ and the size distribution in the tree statistics. Therefore, the moments of the size distribution at the population level must be determined altogether with the population growth rate.

Progress is made when considering the size distribution in the forward lineages. Indeed, integrating out the single-cell growth rate and the division size in \cref{eq_BC_fw} gives: 
\begin{equation}
	\label{eq_rho_sb_fw_exp}
	\rho_{\mathrm{fw}}(s_b)= \int \di s_d' \di s_b' \  \rho_{\mathrm{fw}}(s_b') \kappa(s_b|s_d') \varphi(s_d'- a s_b') \,.
\end{equation}
This self-consistent equation for the forward distribution of newborn sizes $\rho_{\mathrm{fw}}(s_b)$ is independent of $\nu(\alpha)$, which shows that $\rho_{\mathrm{fw}}(s_b)$ is unaffected by fluctuations in single-cell growth rates. The coefficient of variation of newborn sizes in the forward statistics is therefore the same as the one derived in \cite{thomas_analysis_2018} in the absence of noise in single-cell growth rates:
\begin{equation}
	\label{eq_cv_fw}
	\mathrm{CV}_{\mathrm{fw}}^2[s_b] \approx \frac{2-a}{2+a} \mathrm{CV}_{\varphi}^2 + \frac{4}{4-a^2} \mathrm{CV}_{p}^2 \,.
\end{equation}
In the absence of noises in cell size control and volume partitioning, the birth size distribution along forward lineages thus reduces to a Dirac delta function at the mean value $\langle s_b \rangle_{\mathrm{fw}}$. 
When the birth size is the same for all cells along all forward lineages then it is the same for all cells in the tree. Therefore, deviations of the average birth size in the tree statistics from its nominal value $\langle s_b \rangle_{\mathrm{fw}}$ due to fluctuating single-cell growth rates must couple to noises in cell size control and volume partitioning:
\begin{equation}
		\label{eq_sb_exp_coeffs}
    \frac{\langle s_b \rangle_{\mathrm{tree}}}{\langle s_b \rangle_{\mathrm{fw}}} \approx 1 + D_{\varphi} \mathrm{CV}_{\varphi}^2 + D_{p} \mathrm{CV}_{p}^2 
    + D_{\varphi,\alpha} \mathrm{CV}_{\varphi}^2 \mathrm{CV}_{\alpha}^2 
    + D_{p,\alpha} \mathrm{CV}_{p}^2 \mathrm{CV}_{\alpha}^2 \,.
\end{equation}
Here, $\langle s_b \rangle_{\mathrm{fw}}=b/(2-a)$ is imposed by the conservation of volume, $\langle s_d \rangle_{\mathrm{fw}}=2 \langle s_b \rangle_{\mathrm{fw}}$, together with the noisy linear map \cref{eq_csc}. 

We determine coefficients $C$ and $D$ together by expanding \cref{eq_main_powell} for small noise, see Supplementary Material for details. The average birth size in the tree statistics is given by:
\begin{equation}
	\label{eq_sb_cv_exp}
    \frac{\langle s_b \rangle_{\mathrm{tree}}}{\langle s_b \rangle_{\mathrm{fw}}} \approx 1 -\frac{2-a}{2+a} \mathrm{CV}_{\varphi}^2 + \frac{2a}{4-a^2} \mathrm{CV}_{p}^2 
    + \frac{(2-a) \ln 2}{2(2+a)} \mathrm{CV}_{\varphi}^2 \mathrm{CV}_{\alpha}^2 
    -\frac{a \ln 2}{4-a^2} \mathrm{CV}_{p}^2 \mathrm{CV}_{\alpha}^2 \,.
\end{equation}
The first order terms in $\mathrm{CV}_{\varphi}^2$ and $\mathrm{CV}_{p}^2$ were extensively discussed in \cite{thomas_analysis_2018}, so we focus on the second-order terms involving $\mathrm{CV}_{\alpha}^2$, computed for the first time in this work. While noise in size control reduces the average birth size ($D_{\varphi} =-(2-a)/(2+a) \leq 0$), noise in single-cell growth rate coupled to noise in size control increases the average birth size ($D_{\varphi,\alpha} =(2-a) \ln 2/(2(2+a)) \geq 0$, see \cref{fig:lambda_vs_noise}f). Similarly, for noise in volume partitioning, $D_{p} =2a/(4-a^2) \geq 0$ while $D_{p,\alpha} = -a \ln 2/(4-a^2) \leq 0$. Overall, fluctuations in single-cell growth rates reduce the impact of the other sources of noise on the average birth size. 
We show in \cref{fig:lambda_vs_noise}d,e the mean birth size in the tree statistics when varying the noise in single-cell growth rate for different mechanisms $a$ of cell size control, obtained with agent-based simulations detailed in the Methods section. These are in excellent agreement with theoretical predictions.

The population growth rate is given by:
\begin{equation}
	\label{eq_lam_cv_exp}
    \frac{\lambda}{\langle \alpha \rangle} \approx 1- \lp 1 -\frac{\ln 2}{2} \rp \mathrm{CV}_{\alpha}^2 + \frac{(2-a)^2}{2(2+a)\ln 2} \mathrm{CV}_{\alpha}^2 \mathrm{CV}_{\varphi}^2 
    + \frac{2(2-a)-(2+a)\ln 2}{4(2+a)\ln 2} \mathrm{CV}_{\alpha}^2 \mathrm{CV}_{p}^2 \,.
\end{equation}
Let us discuss three important implications of this result.

First, the population growth rate
depends on the strategy $a$ of cell size control. Both $C_{\alpha,\varphi} = (2-a)^2/(2(2+a)\ln 2)$ and $C_{\alpha,p} = (2(2-a)-(2+a)\ln 2)/(4(2+a)\ln 2)$ are decreasing functions of $a$, as shown in \cref{fig:lambda_vs_noise}c, demonstrating that the stronger the mechanism of cell size control (i.e. the closer to the sizer) the larger the population growth rate. In particular, this implies that populations grow faster with the sizer than with the adder, for fixed noise levels $\mathrm{CV}_{\varphi}$, $\mathrm{CV}_{p}$ and $\mathrm{CV}_{\alpha}$. 

Second, noise in size control is always beneficial for population growth, since $C_{\alpha,\varphi} \geq 0$ for all $a$. For the adder, we remind the reader that $\mathrm{CV}_{\varphi}$ is the CV of added size $\Delta_d=s_d-s_b$ between birth and division. Therefore, fluctuations in added size within the adder model increases the population growth rate, which confirms the numerical observations from \cite{thomas_single-cell_2017}.

Third, noise in volume partitioning can be either beneficial or detrimental for population growth: for $a<a_c$ with $a_c=2(2-\ln 2)/(2+\ln 2)$, it increases $\lambda$ ($C_{\alpha,p} \geq 0$), while for $a>a_c$ it decreases it ($C_{\alpha,p} \leq 0$). The threshold value $a_c \approx 0.97$ is very close to the adder mechanism $a=1$, meaning that the noise in volume partitioning has a vanishing effect on population growth for cells following the adder strategy. 

We show in \cref{fig:lambda_vs_noise}a,b the population growth when varying the noise size control and volume partitioning for different mechanisms $a$ of cell size control. Again, dots obtained in numerical simulations fit the theoretical predictions very well. 

When comparing the population growth generated by the different mechanisms of cell size control $a$, on can also consider the situation where the newborn size distribution has the same variability $\mathrm{CV}_{\mathrm{fw}}[s_b]$ for all mechanisms $a$, instead of a common variability $\mathrm{CV}_{\varphi}$ in size control. To do so, we combine \cref{eq_cv_fw} and \cref{eq_lam_cv_exp}, and we derive
\begin{equation}
    \frac{\lambda}{\langle \alpha \rangle} \approx 1 - \lp 1 -\frac{\ln 2}{2} \rp \mathrm{CV}_{\alpha}^2 - \frac{2+\ln 2}{4\ln 2} \mathrm{CV}_{\alpha}^2 \mathrm{CV}_{p}^2 
    + \frac{2-a}{(2+a)\ln 2} \mathrm{CV}_{\alpha}^2 \mathrm{CV}^2_{\mathrm{fw}}[s_b] \,.
\end{equation}
When constraining the different mechanisms of cell size control $a$ to generate forward distributions of birth size with the same CV, we obtain the following: First, noise in volume partitioning is always detrimental to population growth, and is independent of the strategy of cell size control. Second, noise in birth size is always beneficial for population growth and is a decreasing function of $a$. 
Overall, regardless of what is kept constant in the comparison between mechanisms of cell size control ($\mathrm{CV}_{\varphi}$ or $\mathrm{CV}_{\mathrm{fw}}[s_b]$), the population grows faster with the sizer than for the adder and even faster than with the timer.

\subsection{Non-exponential single-cell growth}

\begin{figure*}
	\includegraphics[width=\linewidth]{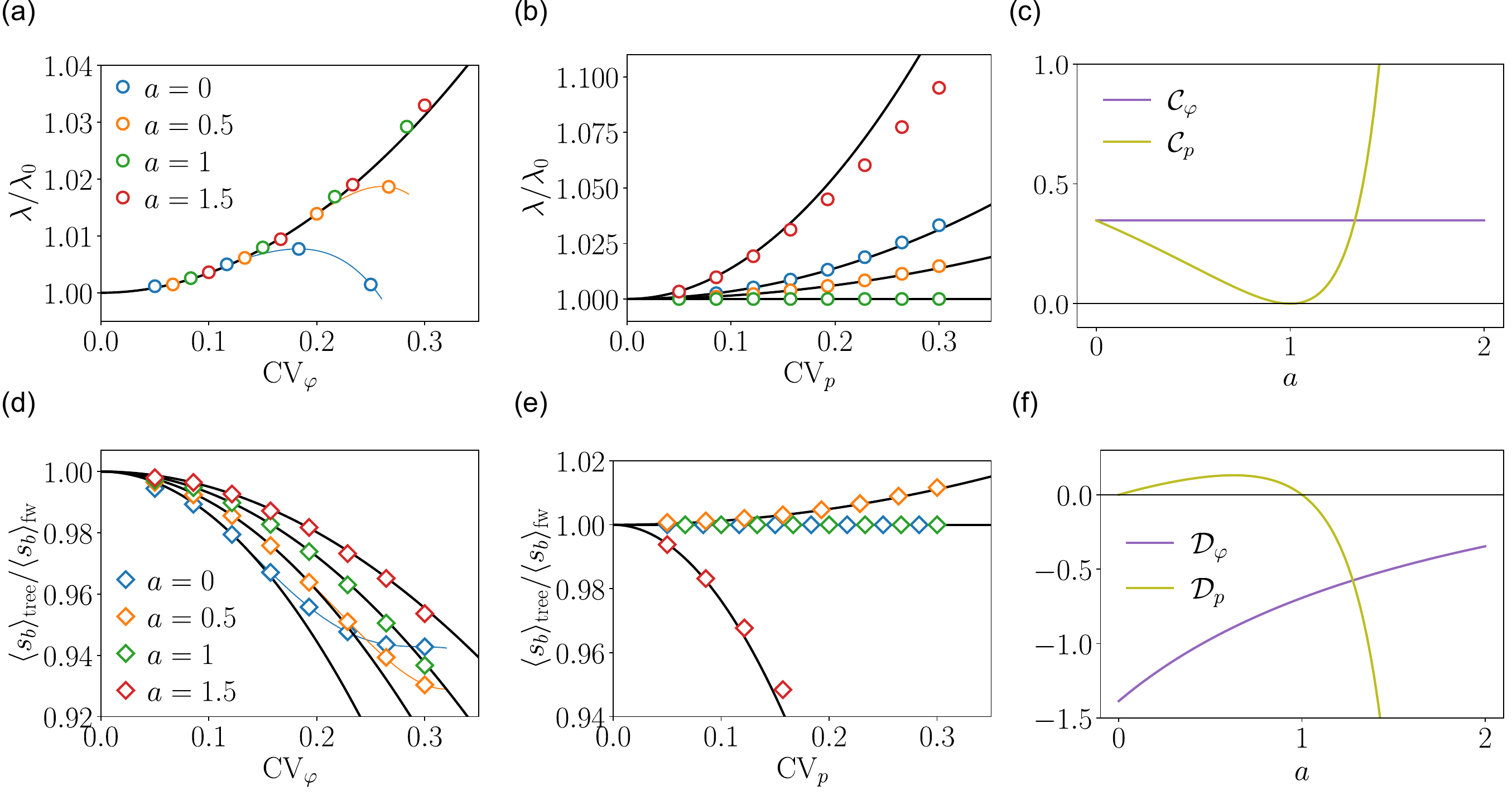}
	\caption{Impact of single-cell variability on population growth and mean birth size for linearly-growing cells. 
		Noises in size control $ \mathrm{CV}_{\varphi}$ and in volume partitioning $\mathrm{CV}_{p}$ are varied, for different models of cell size control: sizer ($a=0$), intermediate between sizer and adder ($a=0.5$), adder ($a=1$) and intermediate between adder and timer, or timer-like, ($a=1.5$). 
		$\lambda$ and $\langle s_b \rangle_{\mathrm{tree}}$ are re-scaled by their values $\lambda_0=\alpha \ln 2/s_b$ and $\langle s_b \rangle_{\mathrm{fw}}=b/(2-a)$ in the absence of noise (all $\mathrm{CV}=0$).
		Theoretical predictions are shown in black and dots and diamonds are the results of numerical simulations. Details on the simulations are explained in the Methods section.
		(a,d) $\mathrm{CV}_{p}=0$. The blue and orange curves are simply indicative of the trend of numerical solutions that deviate from the predictions for large noise and small $a$. (b,e) $\mathrm{CV}_{\varphi}=0$. (a,b,d,e) $\mathrm{CV}_{\alpha}=0$, $ \alpha  =1$, $b=1$. (c,f) Coefficients in the small noise expansions \cref{eq_lambda_lin_coeffs,eq_sb_lin_coeffs} against the mechanism of cell size control $a$.}
	\label{fig_panel_lin_growth}
\end{figure*}

Some cells follow growth patterns which are different from the simple exponential law, such as the bi-linear growth of the fission yeast \cite{horvath_cell_2016}, the super-exponential growth of \textit{E. coli} \cite{kar_distinguishing_2021,genthon_analytical_2022,cylke_super-exponential_2023} and the asymptotic linear growth of \textit{Corynebacterium glutamicum} \cite{messelink_single-cell_2021}, for example. 
Our formalism can be used to derive the population growth rate for these growth laws, which enter via the generation time $\tau_d$ in the generalized Euler-Lotka equation (\ref{eq_main_powell}). We show in the following that, for cells that do not follow pure exponential growth, noise in size control and volume partitioning can provide first order corrections to the population growth rate that are not conditioned to the presence of noise in single-cell growth rates. To illustrate this important result, we consider the simple yet biologically-relevant linear growth, observed recently for \textit{M. tuberculosis} \cite{chung_mycobacterium_2023}. We further show in the Supplementary Material how patterns which depend on the cell cycle progression, like bi-linear growth, can be studied within our framework.

Single-cell linear growth is different from single-cell exponential growth in that, in the latter, the population growth rate is independent of the length-scale of the cells and its nominal value is only given by the mean single-cell growth rate.
This is because the doubling time of a cell (time required to double cell volume), $\tau_2=\tau_d(2 s_b, s_b, \alpha)$, is independent of its size: $\tau_2=\ln(2)/\alpha$. For cells growing linearly on the other hand, whose generation time is given by $\tau_d(s_d,s_b,\alpha)=(s_d-s_b)/\alpha$, the doubling time depends on the birth size as $\tau_2=s_b/\alpha$. Indeed, a big-born cell takes longer to double its volume than a small-born cell if they both accumulate volume at a linear speed $\dot{s}=\alpha$ which is independent of their size. 
Therefore, we expect the presence of corrections to the population growth rate coming from noise in size control and volume partitioning even in the absence of noise in single cell growth rate:
\begin{equation}
	\label{eq_lambda_lin_coeffs}
	\frac{\lambda}{\lambda_0} \approx 1 + \mathcal{C}_{\alpha} \mathrm{CV}_{\alpha}^2 + \mathcal{C}_{\varphi} \mathrm{CV}_{\varphi}^2 + \mathcal{C}_{ p} \mathrm{CV}_{p}^2 \,.
\end{equation}
The nominal value $\lambda_0$ is determined by considering the deterministic dynamics where all cells are born with the same size $s_b$, grow with a unique rate $\alpha$, and divide at the same size $s_d=2 s_b$, so that for linear growth \cref{eq_main_powell} reduces to $1=2 e^{-\lambda s_b/\alpha}$, which implies $\lambda = \alpha \ln 2/s_b$. Therefore, we define $\lambda_0=\langle \alpha \rangle \ln 2 / \langle s_b\rangle_{\mathrm{fw}}$. 

Again, the population growth rate and the mean birth size must be determined together. 
The argument invoked previously, to justify the absence of corrections to the forward average birth size coming from fluctuations in single-cell growth rates that do not couple to noise in cell size control or volume partitioning, is actually independent of the single-cell growth law since \cref{eq_rho_sb_fw_exp} does not involve the generation time $\tau_d$. Therefore, the average birth size in the tree statistics must take the form
\begin{equation}
	\label{eq_sb_lin_coeffs}
	\frac{\langle s_b \rangle_{\mathrm{tree}}}{\langle s_b \rangle_{\mathrm{fw}}} \approx 1 + \mathcal{D}_{\varphi} \mathrm{CV}_{\varphi}^2 + \mathcal{D}_{p} \mathrm{CV}_{p}^2 \,.
\end{equation}
We do not seek the next-order terms involving $\mathrm{CV}_{\alpha}^2$ for $\langle s_b \rangle_{\mathrm{tree}}$ in this case, since $\lambda$ already involves first-order corrections in all sources of noise. 

Solving \cref{eq_main_powell} in the small noise limit allows use to identify the coefficients $\mathcal{C}$ and $\mathcal{D}$. Let us start by discussing the mean birth size in the tree statistics, given by:
\begin{equation}
	\frac{\langle s_b \rangle_{\mathrm{tree}}}{\langle s_b \rangle_{\mathrm{fw}}} \approx 1 - \frac{(4-a) \ln 2}{2+a} \mathrm{CV}_{\varphi}^2 + \frac{4a(1-a)\ln 2}{(2-a)(4-a^2)} \mathrm{CV}_{p}^2 \,.
\end{equation}
Noise in size control always decreases the mean birth size in the tree statistics since $\mathcal{D}_{\varphi} =  -(4-a) \ln 2/(2+a)<0$. On the other hand, noise in volume partitioning with $\mathcal{D}_{p} = 4a(1-a)\ln 2/((2-a)(4-a^2))$ can both increase ($\mathcal{D}_{p}>0$) for $0<a<1$, and decrease ($\mathcal{D}_{p}<0$) for $1<a<2$, the mean birth size in the tree statistics, and has a vanishing effect for the adder and the sizer, as shown in \cref{fig_panel_lin_growth}f. We show in \cref{fig_panel_lin_growth}d,e the mean size when varying $\mathrm{CV}_{\varphi}$ and $\mathrm{CV}_{p}$ for different mechanisms of cell size control. Simulations agree with theoretical predictions very nicely. Note that in \cref{fig_panel_lin_growth}d the smaller $a$ the sooner the numerical results deviate from theoretical predictions, which is because the expansion for a fixed value of $\mathrm{CV}_{\varphi}$ becomes rougher the closer we get to the sizer.

The population growth rate is given by
\begin{equation}
	\label{eq_lam_lin_res}
	\frac{\lambda}{\lambda_0} \approx 1 -\lp 1 -\frac{\ln 2}{2} \rp  \mathrm{CV}_{\alpha}^2 + \frac{\ln 2}{2} \mathrm{CV}_{\varphi}^2 + \frac{2(1-a)^2 \ln 2}{(2-a)^2} \mathrm{CV}_{p}^2 \,.
\end{equation}
The correction coming from fluctuating growth rates is the same as in the case of single-cell exponential growth, $\mathcal{C}_{\alpha}=C_{\alpha}=-(1 -\ln (2)/2)$. 
Noise in size control is always beneficial for population growth ($\mathcal{C}_{\varphi} = \ln(2)/2> 0$), like for single-cell exponential growth, but the corresponding correction is independent of the size control mechanism. 
Noise in volume partitioning is also always beneficial for population growth ($\mathcal{C}_{p}=2(1-a)^2 \ln 2/(2-a)^2 \geq0$), unlike what happens for single-cell exponential growth, and the corresponding correction is non-monotonous with the mechanism of cell size control $a$ and cancels for the adder, as shown in \cref{fig_panel_lin_growth}c. 
Therefore, for linearly-growing cells with fixed noise levels in size control and volume partitioning, the adder is the strategy which leads to the slowest population growth, and the sizer is slower than timer-like strategies down to $a=4/3$. 
We show the results of numerical simulations in \cref{fig_panel_lin_growth}a,b, which agree with theoretical predictions very well.
Moreover, we plot in the Supplementary Material the population growth rate versus the mean birth size in the tree statistics, and show that, contrary to what we intuited with the argument of the cell doubling time, $\lambda$ and $\langle s_b \rangle_{\mathrm{tree}}$ can increase or decrease together when varying the mechanism of cell size control. 

The differences between the cases of exponential and linear growth show the importance of an accurate characterization of the laws of single-cell growth to understand population growth.

\subsection{Cell size control depends on single-cell growth rates}
\label{sec:alpha_lin_map}

\begin{figure*}
	\centering
	\includegraphics[width=\linewidth]{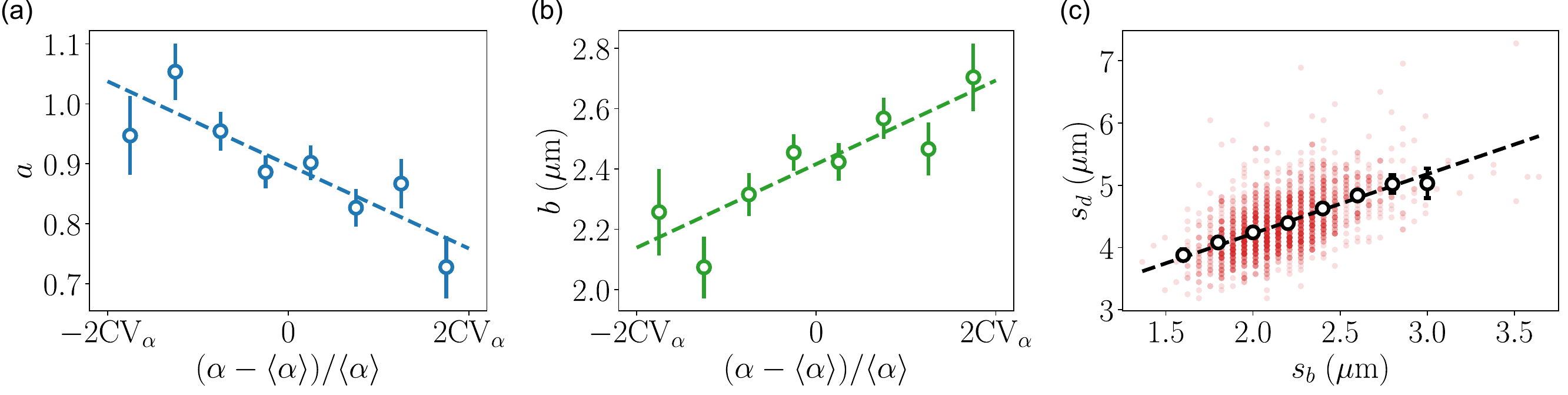}
	\caption{Analysis of mother machine data from Ref. \cite{taheri-araghi_cell-size_2015} for \textit{E. coli} in glucose. Single-cell growth rates in an interval of 4 standard deviations centered around the mean, $(\alpha-\langle \alpha \rangle)/\langle \alpha \rangle \in [- 2\mathrm{CV}_{\alpha}, 2\mathrm{CV}_{\alpha}]$, were divided into 8 bins. For each bin the slope $a$ and intercept $b$ were obtained with a linear fit of $s_d$ vs $s_b$. (a,b)  Error bars represent the uncertainties on the parameters $a$ and $b$ of the best linear fit of $s_d$ vs $s_b$. Best linear fits of $a$ and $b$ vs $(\alpha-\langle \alpha \rangle)/\langle \alpha \rangle$ shown with dashed lines give $\bar{a}=0.90 \pm 0.01$, $S_a = -1.12 \pm 0.25$, $\bar{b}=2.42 \pm 0.03 \SI{}{\micro\meter}$, $S_b=0.89 \pm 0.20$. (c) Example of linear map analysis, for $(\alpha-\langle \alpha \rangle)/\langle \alpha \rangle \in [- \mathrm{CV}_{\alpha},- \mathrm{CV}_{\alpha}/2]$ (third bin in (a,b)). The dashed line indicates the best linear fit of the unbinned data, giving $a=0.95 \pm 0.03$, $b=2.32 \pm 0.07 \mu m$. The black and white circles represent the average size at division for binned size at birth, and are shown for indication only.  
	}
	\label{fig:data_lin_map}
\end{figure*}

Until now, we have assumed that the division size was only determined by the birth size via $\varphi(s_d-a s_b|\alpha) \equiv \varphi(s_d-a s_b)$. To our knowledge, only two theoretical works have relaxed this assumption by considering that, in the context of the adder mechanism, the target added size could depend on the single-cell growth rate \cite{modi_analysis_2017,grilli_empirical_2018}. However, the consequences of this dependence for the population growth rate have not been studied yet. Moreover, it is possible that the mechanism $a$ itself depends on the single-cell growth rate. In the following, we investigate these points for exponentially-growing cells.

We analyze here mother machine data for \textit{E. coli} in different conditions from \cite{taheri-araghi_cell-size_2015}. We binned the data into 8 bins for single-cell growth rates $\alpha$ in an interval of four standard deviations centered around the mean: $(\alpha-\langle \alpha \rangle)/\langle \alpha \rangle \in [- 2\mathrm{CV}_{\alpha}, 2\mathrm{CV}_{\alpha}]$. For each bin in $\alpha$, the linear regression of $s_d$ versus $s_b$ of the unbinned size data point provides the values $a$ and $b$ of the noisy linear map. An example, for single-cell growth rates $(\alpha-\langle \alpha \rangle)/\langle \alpha \rangle \in [- \mathrm{CV}_{\alpha},- \mathrm{CV}_{\alpha}/2]$ in glucose is shown in \cref{fig:data_lin_map}c. Here, the vertical stripes are not due to binning of the size data points but comes from the resolution of the size measurements. The black dashed line shows the best linear fit of the unbinned size data points in red. The black and white dots are the average sizes at division when binning the size at birth, they are not used for the fit and are simply shown for visual confirmation that the trend is indeed linear and follows the best fit. 
The values of $a$ and $b$ for all 8 bins are reported in \cref{fig:data_lin_map}a,b for the glucose condition. This reveals that both the slope $a$ and intercept $b$ of the linear map depend on the single-cell growth rate. Fast-growing cells tend to follow an intermediate sizer-adder strategy while slow-growing cells tend to follow a near-adder strategy. The same analysis is conducted for \textit{E. coli} in Sorbitol and Tryptic Soy Broth and the results are shown in the Supplementary Material.

This analysis encouraged us to define the slope and intercept of the linear map as:
\begin{align}
	a &= \Bar{a} \lp 1 + S_a \ \frac{\alpha-\langle \alpha \rangle}{\langle \alpha \rangle} \rp \\
	b &= \Bar{b} \lp 1 + S_b \ \frac{\alpha-\langle \alpha \rangle}{\langle \alpha \rangle} \rp \,,
\end{align}
where $\Bar{a}= \int \di \alpha \ \nu(\alpha) a$ and $\Bar{b}= \int \di \alpha \ \nu(\alpha) b$ are the average slope and intercept, and where $S_a$ and $S_b$ are the sensitivities of the slope and intercept on the single-cell growth rate. For example, the analysis in glucose gives $\bar{a}=0.90 \pm 0.01$, $S_a = -1.12 \pm 0.25$, $\bar{b}=2.42 \pm 0.03 \SI{}{\micro\meter}$ and $S_b=0.89 \pm 0.20$. For simplicity we suppose that the noise $\eta$ remains independent of the single-cell growth rate.

A key difference with the cases treated until now is that the forward statistics of birth sizes explicitly depends on fluctuations in single-cell growth rates, as observed previously in \cite{modi_analysis_2017}. This can be seen with the boundary condition \cref{eq_BC_fw} integrated over $s_d$ and $\alpha$:
\begin{equation}
	\label{eq_self_cons_fw_alpha_dep}
	\rho_{\mathrm{fw}}(s_b)=  \int \di s_d' \di s_b' \di \alpha' \  \rho_{\mathrm{fw}}(s_b') \nu(\alpha') \kappa(s_b|s_d') \varphi(s_d'- a s_b'|\alpha') \,,
\end{equation}
which involves the distribution of single-cell growth rates. 
Using $\langle s_b \rangle_{\mathrm{fw}}=\bar{b}/(2-\bar{a})$,
we multiply \cref{eq_self_cons_fw_alpha_dep} by $s_b^2$ and solve the equation in the small noise limit which gives:
\begin{equation}
	\mathrm{CV}_{\mathrm{fw}}^2[s_b] \approx  \frac{2-\Bar{a}}{2+\Bar{a}} \mathrm{CV}_{\varphi}^2  + \frac{4}{4-\Bar{a}^2} \mathrm{CV}_{p}^2 
	+ \frac{(\Bar{a}S_a +(2-\Bar{a}) S_b)^2}{4-\Bar{a}^2} \mathrm{CV}_{\alpha}^2 \,,
\end{equation}
where $ \mathrm{CV}_{\varphi}^2=\langle \eta^2 \rangle_{\mathrm{fw}} /\bar{b}^2$.
Therefore, unlike in the previous cases, fluctuations in single-cell growth rates will directly impact the mean birth size in the tree statistics with a first order correction which is not conditioned on the presence of noises in size control or volume partitioning:
\begin{equation}
	\frac{\langle s_b \rangle_{\mathrm{tree}}}{\langle s_b \rangle_{\mathrm{fw}}} \approx 1 + d_{\varphi} \mathrm{CV}_{\varphi}^2 + d_{p} \mathrm{CV}_{p}^2 + d_{\alpha}  \mathrm{CV}_{\alpha}^2 \,.
\end{equation}
The first-order change in mean birth size induced by fluctuations in single-cell growth rates will feedback on the population growth rate since they are coupled by the Euler-Lotka equation (\ref{eq_main_powell}). As a consequence we expect the first-order correction to the population growth rate coming from fluctuations in single-cell growth rates to be different from $-(1-\ln(2)/2) \mathrm{CV}_{\alpha}^2$ derived previously in \cref{eq_lam_cv_exp,eq_lam_lin_res}. We thus focus on the leading order term:
\begin{equation}
	\frac{\lambda}{\langle \alpha \rangle} \approx 1 + c_{\alpha} \mathrm{CV}_{\alpha}^2 \,.
\end{equation}
Solving the Euler-Lotka equation in the small noise limit provides the expressions of coefficients $c$ and $d$. The mean birth size reads:
\begin{equation}
	\frac{\langle s_b \rangle_{\mathrm{tree}}}{\langle s_b \rangle_{\mathrm{fw}}} \approx 1 -\frac{2-\Bar{a}}{2+\Bar{a}} \mathrm{CV}_{\varphi}^2 + \frac{2\Bar{a}}{4-\Bar{a}^2} \mathrm{CV}_{p}^2 
	+ \frac{(\Bar{a}S_a + (2-\Bar{a}) S_b)((2+\Bar{a})\ln 2 - \Bar{a}S_a - (2-\Bar{a}) S_b)}{4-\Bar{a}^2} \mathrm{CV}_{\alpha}^2 \,.
\end{equation}
The new correction coming from fluctuations in single-cell growth rates can both increase or decrease the mean birth size depending on the average mechanism of cell size control $\bar{a}$ and the specificities $S_a$ and $S_b$. 

\begin{figure*}
	\centering
	\includegraphics[width=\linewidth]{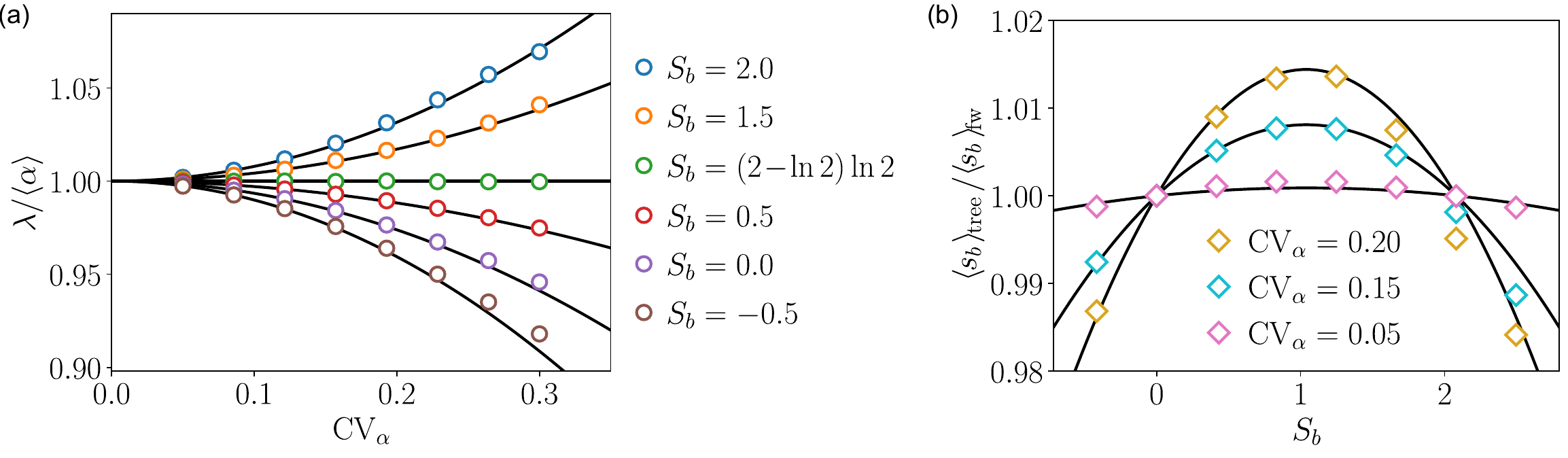}
	\caption{Impact of single-cell variability on population growth and mean birth size with $\alpha$-dependent cell size control. (a) Population growth rate and (b) mean birth size in the tree statistics, for the adder mechanism ($\bar{a}=1$,  $S_a=0$, $\bar{b}=1$), in the absence of noise in size control and size partitioning ($\mathrm{CV}_{\varphi}=\mathrm{CV}_{p}=0$), for various values of the target added size sensitivity on single-cell growth rate, $S_b=S_{\Delta}$, and for various noise levels in single cell growth rates $\mathrm{CV}_{\alpha}$. The dots and diamonds are the results of numerical simulations detailed in the Methods section and the theoretical predictions are shown in black.}
	\label{fig:lambda_vs_noise_alpha_dep_lin_map}
\end{figure*}

The population growth rate is given by:
\begin{equation}
	\label{eq_lam_alp_lin_map}
	\frac{\lambda}{\langle \alpha \rangle} \approx 1 -\lp 1 - \frac{\ln 2}{2} - \frac{\Bar{a}S_a + (2-\Bar{a}) S_b}{2 \ln 2} \rp \mathrm{CV}_{\alpha}^2 \,.
\end{equation}
Unlike what happens when the target division size does not depend on the single-cell growth rate, here the population growth rate explicitly depends on the mechanism of cell size control, even in the absence of noise in size control or volume partitioning. 
The sign of $S_a-S_b$ controls the behavior of the population growth rate as a function of the average mechanism of size control $\Bar{a}$. If $S_a-S_b<0$, $\lambda$ decreases with $\Bar{a}$ and the sizer leads to faster population growth than the adder. If $S_a-S_b>0$, the opposite is true and the adder leads to faster population growth than the sizer, but is slower than timer-like strategies. 
Importantly, for large values of the sensitivities $S_a$ and/or $S_b$, that is when slow-growing cells divide at smaller sizes than fast-growing cells, population growth is enhanced by variability in single-cell growth rates ($c_{\alpha}>0$) for all average mechanisms of cell size control $\bar{a}$. 

Let us illustrate this result with the example of an adder, $\Bar{a}=1$ and $S_a=0$, where the target added size $b=\Delta$ depends on the single-cell growth rate, as considered in \cite{modi_analysis_2017,grilli_empirical_2018}. Note that for the adder $\Bar{a}=1$, $\lambda$ depends only on $S_a+S_b$, so $S_a$ and $S_b$ play symmetric roles and similar results are obtained when keeping $S_b=0$ and varying $S_a$ instead. 
We show in \cref{fig:lambda_vs_noise_alpha_dep_lin_map}a the population growth rate as a function of $\mathrm{CV}_{\alpha}$ for different values of the sensitivity of the target added size on the single-cell growth rate, $S_b=S_{\Delta}$, and in \cref{fig:lambda_vs_noise_alpha_dep_lin_map}b the mean birth size as a function of $S_b$ for different values of $\mathrm{CV}_{\alpha}$.
When $S_b=0$ (purple dots), we recover the case treated in the first part of the article.
For $S_b=(2-\ln 2)\ln 2$ (green dots), $\lambda$ is independent of the noise in single-cell growth rates.
Above this value, fast-growing cells grow by an average added size which is sufficiently larger than that of the slow-growing cells, so that fluctuations in single-cell growth rates become beneficial for population growth.

This analysis also shows that the bounds on the population growth rate, $\langle \alpha^{-1} \rangle^{-1} \leq \lambda \leq \langle \alpha \rangle$, derived in \cite{thomas_single-cell_2017} when $\psi(s_d|s_b,\alpha)=\psi(s_d|s_b)$ are no longer valid in this case. We show in the Supplementary Material that, on the other hand, the lower bound $\ln 2/\langle \tau_d \rangle_{\mathrm{fw}} \leq \lambda$ derived in \cite{genthon_fluctuation_2020} (see \cite{genthon_fluctuations_2022}, p.35 for details) holds.

\section{Discussion}

In this article we derived a generalized Euler-Lotka equation that connects the population growth rate to the distribution of birth sizes, division sizes and single-cell growth rates in the population statistics. 
While in previous works only the role of the variability in single-cell growth rates has been investigated, 
we obtained analytical expressions for the population growth rate and for the mean birth size in the presence of fluctuations in cell size control, volume partitioning upon division and single-cell growth rates.

We illustrated our findings in 
three
biologically-relevant cases:
(i) For exponentially-growing cells, 
population growth is reduced by fluctuations in single-cell growth rates but this reduction is almost always attenuated by fluctuations in size control and volume partitioning.
Importantly, the population growth rate depends on the mechanism of cell size control, which allows for a comparison between 
the different mechanisms.
(ii) For linearly-growing cells, like \textit{M. tuberculosis} for example \cite{chung_mycobacterium_2023}, 
noisy size control and noisy volume partitioning impact the population growth rate even in the absence of noise in single-cell growth rates. These corrections are always beneficial for population growth. 
The analysis also revealed the complex non-monotonous relation between the population growth rate and the mean birth size in the population.
(iii) When cell size control depends on the single-cell growth rate, which was observed when analysing \textit{E. coli} data from \cite{taheri-araghi_cell-size_2015}, fluctuations in single-cell growth rates can be both detrimental or beneficial for population growth. For example, these fluctuations are beneficial within the adder model if the added volume is strongly positively correlated to the single-cell growth rate.

The method proposed in the article to determine the dependence of the population growth rate on single-cell variability is very general. 
For example, our approach equally applies to asymmetrically-dividing cells like budding yeast \textit{S. cerevisiae}. 
We show in the Supplementary Material how to recover the expression of the population growth rate as a function of the division asymmetry for the sizer mechanism, first obtained in \cite{barber_modeling_2021}. There, the volume partitioning is asymmetric but noiseless and there is no noise in cell size control either, but these assumptions could be relaxed in our framework. 
More generally, it will be interesting to incorporate additional fluctuating variables which are relevant for growth and which are asymmetrically-segregated upon division, like damage proteins that affect single-cell growth rates \cite{lin_optimal_2019} or parasite infections that modulate the division rate \cite{marguet_spread_2024}. 

We hope that our findings will lead to valuable contributions in different research areas, of which we give three examples. 

First, observations of individual cancer cells are becoming available which allows for a characterization of their growth pattern \cite{cermak_high-throughput_2016,mu_mass_2020}. Moreover, the size control of cancer cells is now investigated \cite{miotto_size-dependent_2024}. Our results may thus provide insights for a better understanding of cancer development based on these newly-available data.

Second, the recent development of micro-fluidic devices where single-lineages (forward lineages) of cells are grown, like the mother machine \cite{wang_robust_2010}, has significantly increased the amount of available data.
A major challenge lies in the inference of population properties, such as the population growth rate, from single-lineage data \cite{genthon_fluctuation_2020,levien_large_2020,pigolotti_generalized_2021}. In our framework, the noise levels $ \mathrm{CV}_{\alpha}$, $\mathrm{CV}_{\varphi}$ and $\mathrm{CV}_{p}$ are evaluated in the single-lineage statistics, thus our findings provide estimations of the population growth rate from fluctuations measured in mother machines. 

Third, understanding why the adder mechanism is so ubiquitous across different domains of life remains an important question in cell biology. This question has recently been reshaped as an optimization problem where the optimal mechanism of cell size control is the one maximizing the population growth rate (population fitness) in the presence of cell death \cite{hobson-gutierrez_evolutionary_2023}, or minimizing the spread of the birth size distribution \cite{proulx-giraldeau_evolution_2022,elgamel_effects_2024}. Our results quantify the impact of the size control mechanism (characterized by the parameter $a$) on both the population growth rate and the size statistics, which sheds some light on the evolutionary advantages of cell size control.

\section{Methods}

\subsection{Numerical simulation details}
\label{app_num}

We simulate populations of cells starting from one ancestor cell, and let them grow until they reach $N_f=5 \times 10^6$ unless otherwise stated. The single-cell growth rates are drawn from a log-normal distribution with mean $\langle \alpha \rangle=1$ and variance $\sigma_{\alpha}^2$. The sizes at division for cells born with size $s_b$ are drawn from a Gaussian distribution with mean $\langle s_d | s_b,\alpha \rangle_{\mathrm{fw}} = a s_b +b$ and variance $\langle \eta^2 \rangle_{\mathrm{fw}} = b^2 \mathrm{CV}_{\varphi}^2$. The daughter-to-mother volume ratio $p$ for one of the two daughter cells is drawn from a symmetric Beta distribution $\mathrm{Beta}(\beta,\beta)$ with $\beta=1/(8 \sigma_p^2)-1/2$, where $\langle p \rangle=1/2$ and where $\sigma_p^2$ is the variance of the volume fraction $p$. The fraction of volume inherited by the other daughter cell is $1-p$.

The population growth rate is computed using the data in the final time window $\Delta t=5$. We fit both $\ln N$ and $\ln V$ versus $t$ by a linear function, with $V$ the total volume of the population, and check that the two slopes are equal as expected with cell size control. The value of the slope is used as the population growth rate. 

In \cref{fig_panel_lin_growth}a,b,d,e the algorithm is run once only for each condition.
In \cref{fig:lambda_vs_noise}a,b,d,e the algorithm is run $m=20$ times for each condition and the error bars represent the two-sided $95\%$ confidence interval $\mu_m \pm t_m \sigma_m/\sqrt{m}$  with $\mu_m$ and $\sigma_m$ the sample mean and sample standard deviation, and $t_m$ the Student coefficient with $t_{20}=2.093$.
In \cref{fig:lambda_vs_noise}d, we used $N_f=10^7$ for all four values of $a$, and ran the algorithm $m=200$ times for $a=1.5$ ($t_{200} \approx 1.984$). Note that the error bars are really small and only visible in \cref{fig:lambda_vs_noise}d,e for $a=1.5$.

Given the small amplitude of the corrections in \cref{fig:lambda_vs_noise}a,b,d,e, we need to keep terms up to `total' order $\mathrm{CV}^4$ for all sources of noise for better precision. For the population growth rate in \cref{fig:lambda_vs_noise}a,b, this means keeping terms in $\mathrm{CV}_{\alpha}^3$ and $\mathrm{CV}_{\alpha}^4$, which are obtained by expanding the simple Euler-Lotka equation $1= \int \di \alpha \ 2^{1-\lambda/\alpha} \nu(\alpha)$ in the absence of noise other than in single-cell growth rates: 
\begin{align}
	\frac{\lambda^*}{\langle \alpha \rangle} \approx 1 &- \lbk 1 -\frac{\ln 2}{2} \rbk \mathrm{CV}_{\alpha}^2 + \lbk 1 -\ln 2 + \frac{(\ln 2)^2}{6} \rbk  \gamma_{\alpha} \mathrm{CV}_{\alpha}^3  \nn \\
	&+  \Bigg[  1- 2 \ln 2 + (\ln 2)^2 - \frac{(\ln 2)^3}{8}
	+ k_{\alpha} \lp -1 + \frac{3 \ln 2}{2} - \frac{(\ln 2)^2}{2} + \frac{(\ln 2)^3}{24}\rp \Bigg] \mathrm{CV}_{\alpha}^4 \,,
\end{align}
where $ \gamma_{\alpha}$ and $ k_{\alpha}$ are the skewness and the kurtosis of the single-cell growth rate distribution $\nu$. 

Similarly, for the mean birth size in \cref{fig:lambda_vs_noise}d,e we keep corrections coming from noise in size control and volume partitioning up to `total' order 4. Using the method from \cite{thomas_analysis_2018}, we get:
\begin{align}
	\frac{\langle s_b \rangle_{\mathrm{tree}}^*}{\langle s_b \rangle_{\mathrm{fw}}} \approx &1 -\frac{2-a}{2+a} \mathrm{CV}_{\varphi}^2 + \frac{2a}{4-a^2} \mathrm{CV}_{p}^2 + \frac{(2-a)^2 \gamma_{\varphi}}{4+2a+a^2}  \mathrm{CV}_{\varphi}^3
	+ \frac{2 (480 + 292 a + 86 a^2 + 13 a^3) }{(2 + a)^2 (4 + 2 a + a^2)} \mathrm{CV}_{\varphi}^2 \mathrm{CV}_{p}^2 \nn \\
	& + \frac{(2 - a)^2 (4 - 5 a^2 - (4 - a^2) k_{\varphi})}{(2 + a)^2 (4 + a^2)} \mathrm{CV}_{\varphi}^4 +\frac{ 64 - 8 a^2 + 28 a^3 - 6 a^4 + a^5 + 4a(4  - a^2 )k_p}{(2 - a)^2 (2 + a)^2 (4 + a^2) (4 + 2 a + a^2)}  2 a^2 \mathrm{CV}_{p}^4 \,,
\end{align}
with $ \gamma_{\varphi}$ and $ k_{\varphi}$ the skewness and the kurtosis of distribution $\varphi$, and $k_p$ the kurtosis of the volume partitioning distribution $\pi$. This expansion does not involve any term in $\mathrm{CV}_{p}^3$ since distribution $\pi$ is symmetric.

\section*{Acknowledgments}

I am indebted to Jie Lin and Gabin Laurent for their help with the simulations. I warmly thank Philipp Thomas for stimulating discussions, and David Lacoste, Stephan Grill and Pierre Haas for their valuable feedbacks on the manuscript.

\end{document}


\title{Supplementary Material for \\`From noisy cell size control to population growth: when variability can be beneficial'}
\author{Arthur Genthon}
\affiliation{Max Planck Institute for the Physics of Complex Systems, 01187 Dresden, Germany}

\maketitle

\renewcommand{\theequation}{S\arabic{equation}}
\setcounter{equation}{0}

\renewcommand{\thefigure}{S\arabic{figure}}
\setcounter{figure}{0}

\tableofcontents

\section{Derivation of the generalized Euler-Lotka equation with continuous rate models}

In the literature, population dynamics is often modeled by an evolution equation in which the division size $s_d$ is not an explicit variable. Instead, in these models cells divide with a division rate per unit time $r(\tau,s_b,\alpha)$ which can depend on age $\tau$, single-cell growth rate $\alpha$, absolute size $s(\tau,s_b,\alpha)$ and added size $s-s_b$, for example \cite{robert_division_2014,osella_concerted_2014,taheri-araghi_cell-size_2015,thomas_single-cell_2017,genthon_analytical_2022}.
In this section, we show that the two approaches are equivalent by deriving the generalized Euler-Lotka equation with a division rate formalism.

In a freely-growing population, the expected number of cells $n(\tau,s_b,\alpha;t)$ with age $\tau$, birth size $s_b$ and growth rate $\alpha$ in the population at time $t$ follows:
\begin{equation}
	\label{eq:pbe_alpha_CRM}
	\partial_t n(\tau,s_b,\alpha;t) = - \partial_{\tau}  n(\tau,s_b,\alpha;t) - r(\tau,s_b,\alpha) n(\tau,s_b,\alpha;t) \,,
\end{equation}
where $r(\tau,s_b,\alpha)$ is the division rate per unit time. 
The boundary condition for newborn cells reads:
\begin{equation}
	\label{eq:bc_alpha_CRM}
	n(\tau=0,s_b,\alpha;t) = 2 \int \di \tau' \di s_b' \di \alpha' \ 
	\Sigma(s_b,\alpha|\tau',s_b',\alpha') r(\tau',s_b',\alpha') n(\tau',s_b',\alpha';t) \,,
\end{equation}
where $\Sigma(s_b,\alpha|\tau',s_b',\alpha')$ is the probability for a cell to be born with birth size $s_b$ and growth rate $\alpha$ if its mother was born with birth size $s_b'$, growth rate $\alpha'$ and divided at age $\tau'$. It is related to the kernel $\mathcal{K}$ introduced in the main text by $\mathcal{K}(s_d,s_b,\alpha|s_d',s_b',\alpha')=\psi(s_d|s_b,\alpha)\Sigma(s_b,\alpha|\tau_d(s_d',s_b',\alpha'),s_b',\alpha')$, with $\tau_d$ the generation time.
In the steady-state limit, the population grows like:
\begin{equation}
	\label{eq:n_ss_CRM}
	n(\tau,s_b,\alpha;t) \underset{t \to \infty}{\sim} \Pi(\tau,s_b,\alpha) e^{\lambda t} \,,
\end{equation}
where $\Pi(\tau,s_b,\alpha)$ is the joint snapshot distribution of age, birth size and single-cell growth rate, and $\lambda$ the population growth rate.
When plugging \cref{eq:n_ss_CRM} into \cref{eq:pbe_alpha_CRM}, we solve the equation for the joint distribution in steady-state:
\begin{equation}
	\label{eq_Pi_sol_CRM}
	\Pi(\tau,s_b,\alpha)=\Pi(0,s_b,\alpha) e^{-\lambda \tau - \int_{0}^{\tau} \di a \ r(a,s_b,\alpha)} \,.
\end{equation}
We then define the joint distribution of birth size and single-cell growth rate for newborn cells in the tree statistics:
\begin{equation}
	\label{eq_rho_def_CRM}
	\rho_{\mathrm{tree}}(s_b,\alpha)= \frac{\Pi(0 ,s_b,\alpha)}{\int \di s_b' \di \alpha' \Pi(0 ,s_b',\alpha')} \,.
\end{equation}
Combining \cref{eq:n_ss_CRM,eq_Pi_sol_CRM,eq_rho_def_CRM} into \cref{eq:bc_alpha_CRM} leads to:
\begin{equation}
	\rho_{\mathrm{tree}}(s_b,\alpha) = 2 \int \di \tau' \di s_b' \di \alpha' \ 
	\Sigma(s_b,\alpha|\tau',s_b',\alpha') r(\tau',s_b',\alpha') e^{ - \int_{0}^{\tau'} \di a \ r(a,s_b,\alpha)}  \rho_{\mathrm{tree}}(s_b',\alpha') e^{-\lambda \tau'} \,.
\end{equation}
This equation depends on the division rate only via the forward conditional distribution of generation times
\begin{equation}
	\label{eq_phi_CRM}
	\phi_{\mathrm{fw}}(\tau_d|s_b,\alpha) =r(\tau_d,s_b,\alpha) e^{-\int_{0}^{\tau_d} r(\tau,s_b,\alpha) d \tau} \,,
\end{equation}
where $\exp[-\int_{0}^{\tau_d} r(\tau,s_b,\alpha) d \tau]$ is the probability for a cell born with size $s_b$ and growth rate $\alpha$ not to divide up to age $\tau_d$, and $r(\tau_d,s_b,\alpha) d\tau_d$ is the probability for this cell to divide with age in $[\tau_d;\tau_d + d\tau_d]$. 
A change of variable $\psi_{\mathrm{fw}}(s_d|s_b,\alpha) d s_d= \phi_{\mathrm{fw}}(\tau_d|s_b,\alpha) d \tau_d$, where $\psi_{\mathrm{fw}}(s_d|s_b,\alpha)$ is the conditional distribution of division sizes, gives
\begin{equation}
	\rho_{\mathrm{tree}}(s_b,\alpha) = 2 \int \di s_d' \di s_b' \di \alpha' \ 
	\hat{\Sigma}(s_b,\alpha|s_d',s_b',\alpha')  \psi_{\mathrm{fw}}(s_d'|s_b',\alpha') \rho_{\mathrm{tree}}(s_b',\alpha') e^{-\lambda \tau_d(s_d',s_b',\alpha')}\,,
\end{equation}
with $\hat{\Sigma}(s_b,\alpha|s_d',s_b',\alpha') = 	\Sigma(s_b,\alpha|\tau_d(s_d',s_b',\alpha'),s_b',\alpha') $.
Like in the main text, we focus on a class of kernels $	\hat{\Sigma}(s_b,\alpha|s_d',s_b',\alpha')=\nu(\alpha)\kappa(s_b|s_d')$, which implies that the size at birth and the single-cell growth rate are independent random variables in the tree statistics: $\rho_{\mathrm{tree}}(s_b,\alpha)=\rho_{\mathrm{tree}}(s_b)\rho_{\mathrm{tree}}(\alpha)$ with $\rho_{\mathrm{tree}}(\alpha)=\nu(\alpha)$. 

Finally, when integrating $\alpha$, we recover 
\begin{equation}
	\rho_{\mathrm{tree}}(s_b) = 2 \int \di s_d' \di s_b' \di \alpha' \ 
	\kappa(s_b|s_d') \psi_{\mathrm{fw}}(s_d'|s_b',\alpha')  \rho_{\mathrm{tree}}(s_b') \nu(\alpha') e^{-\lambda \tau_d(s_d',s_b',\alpha')}\,.
\end{equation}
Since the division rate does not appear explicitly in the result, but is instead only involved in distribution $\psi_{\mathrm{fw}}(s_d|s_b,\alpha)$, we choose to present a mathematically simpler yet equivalent version in the main text where $s_d$ is treated as a variable and $\psi_{\mathrm{fw}}(s_d|s_b,\alpha)$ is an input.

\section{Cell size control in noisy linear maps and continuous rate models}

The definitions of the adder in the noisy linear map and in the continuous rate models are equivalent, and the definitions of the other mechanisms of cell size control are consistent up to a small noise approximation.
To prove this, it is useful to define the division rate per unit size $h(s(\tau,s_b,\alpha),s_b,\alpha)=r(\tau,s_b,\alpha)/g(\tau,s_b,\alpha)$, where $g(\tau,s_b,\alpha)=\di s /\di \tau$ is the single-cell growth speed. 
In continuous rate models, the conditioned distribution of division sizes is obtained from \cref{eq_phi_CRM} and reads
\begin{equation}
	\label{eq_CRM_phi}
	\psi_{\mathrm{fw}}(s_d|s_b,\alpha) = h(s_d,s_b,\alpha) e^{-\int_{s_b}^{s_d} h(s,s_b,\alpha) d s} \,.
\end{equation}

The adder model is defined by a division rate per unit size $h(s,s_b,\alpha) \equiv h(s-s_b,\alpha)$ which depends on added size but neither on size nor age \cite{taheri-araghi_cell-size_2015}, which gives $\psi_{\mathrm{fw}}(s_d|s_b,\alpha) =h(s_d-s_b,\alpha) \exp[-\int_{0}^{s_d-s_b} h(\Delta,\alpha) d\Delta]$.
This shows that the definition of the adder in continuous rate models is consistent with the definition of the stochastic linear map, since $\psi_{\mathrm{fw}}(s_d|s_b,\alpha) = \varphi_{\mathrm{fw}}(s_d-s_b|\alpha)$ depends only on $s_d-s_b$.

The sizer model is defined by a division rate per unit size $h(s,s_b,\alpha) \equiv h(s,\alpha)$ that depends on size, but neither on added size nor age \cite{taheri-araghi_cell-size_2015}, which gives $\psi_{\mathrm{fw}}(s_d|s_b,\alpha) = h(s_d,\alpha) \exp[-\int_{s_b}^{s_d} h(s,\alpha) ds]$.
In general, $\psi_{\mathrm{fw}}(s_d|s_b,\alpha)$ is not independent of the size at birth $s_b$. However, in the small noise limit when the distributions of birth sizes and division sizes are peaked, the division rate $h$ is vanishing for typical values of birth size $s_b$, which allows the approximation $s_b=0$ for the lower bound the integral \cite{taheri-araghi_cell-size_2015}. In this case, the definition of the sizer in continuous rate models is consistent with the definition of the stochastic linear map, where $\psi_{\mathrm{fw}}(s_d|s_b,\alpha)=\varphi_{\mathrm{fw}}(s_d|\alpha)$ depends only on $s_d$ and is independent of $s_b$.

The other mechanisms of cell size control, characterized by other values of the slope $a$ of the noisy linear map, are obtained in continuous rate models with division rates per unit size of the form $h(s-a s_b,\alpha)$, up to a small noise approximation similar to that used for the sizer model.

\section{Mathematical derivation of the coefficients in the small noise expansions}
\label{app_deriv}

In this section, we show how to derive the analytical expressions for the population growth rate and the mean birth size in the tree statistics, eqs. (16) and (17) in the main text, from the generalized Euler-Lotka equation eq. (6). The solutions of the other cases, namely with single-cell linear growth and single-cell-growth-rate-dependent target division size, follow the same steps. We start by integrating out $s_d$ and $\alpha$ in eq. (6) from the main text:
\begin{equation}
	\label{eq_powell_sb}
	\rho_{\mathrm{tree}}(s_b) = 2 \int \di s_d' \di s_b' \di \alpha' \ 
	\kappa(s_b|s_d') \rho_{\mathrm{tree}}(s_b') \nu(\alpha')\varphi(s_d'-a s_b')\lp \frac{s_b'}{s_d'} \rp^{\lambda / \alpha'}\,.
\end{equation}
We next use this equation to compute the first moments of the distribution of birth sizes, $\langle s_b^n \rangle_{\mathrm{tree}}$ for $n=0, 1, 2$. 

Let us start with $n=0$: direct integration of \cref{eq_powell_sb} together with the normalization of the volume partitioning kernel $\int \di s_b \ \kappa(s_b|s_d') =1$ gives
\begin{equation}
	\label{eq_self_cons_1}
	1 = 2 \int \di s_d' \di s_b' \di \alpha' \ 
	\rho_{\mathrm{tree}}(s_b') \nu(\alpha') \varphi(s_d'-a s_b') \lp \frac{s_b'}{s_d'} \rp^{\lambda / \alpha'}\,.
\end{equation}
The population growth rate $\lambda$ implicitly depends on the newborn size distribution $\rho_{\mathrm{tree}}$, the single-cell growth rate distribution $\nu$ and the size control distribution $\varphi$, but not on the size partitioning kernel $\kappa$, we therefore first seek the results in the form: 
\begin{equation}
	\label{eq_lam_exp_app}
	\frac{\lambda}{\langle \alpha \rangle} \approx 1 + \hat{C}_{\alpha} \mathrm{CV}_{\alpha}^2 + \hat{C}_{\alpha, \varphi} \mathrm{CV}_{\alpha}^2 \mathrm{CV}_{\varphi}^2 + \hat{C}_{\alpha, s_b} \mathrm{CV}_{\alpha}^2 \mathrm{CV}^2_{\mathrm{tree}}[s_b] \,,
\end{equation}
and
\begin{equation}
	\label{eq_sb_exp_app}
	\frac{\langle s_b \rangle_{\mathrm{tree}}}{\langle s_b \rangle_{\mathrm{fw}}} \approx 1  + \hat{D}_{\varphi} \mathrm{CV}_{\varphi}^2 + \hat{D}_{s_b} \mathrm{CV}^2_{\mathrm{tree}}[s_b] + \hat{D}_{\varphi,\alpha} \mathrm{CV}_{\varphi}^2 \mathrm{CV}_{\alpha}^2 
	+ \hat{D}_{s_b,\alpha} \mathrm{CV}^2_{\mathrm{tree}}[s_b] \mathrm{CV}_{\alpha}^2 \,.
\end{equation}
The CV of the newborn size in the tree statistics, $\mathrm{CV}_{\mathrm{tree}}[s_b]$, will be determined explicitly in a later step. 
Assuming that the distributions $\rho_{\mathrm{tree}}$, $\nu$ and $\varphi$ in \cref{eq_self_cons_1} are peaked, we write $s_b=\langle s_b\rangle_{\mathrm{tree}} + \epsilon_{s_b}$, $\alpha=\langle \alpha\rangle + \epsilon_{\alpha}$ and $s_d-a s_b=b + \epsilon_{s_d}$ and we expand the integrand $f(s_b,s_d,\alpha)=\lp s_b/s_d \rp^{\lambda / \alpha}$ up to order 2 for each variable: $f(x_1,x_2,x_3) \approx
f(\langle x_1\rangle,\langle x_2\rangle,\langle x_3\rangle)  + \sum_{i=1}^{3} \lp  \epsilon_{x_i}^2 \partial_{{x_i}^2} f /2  + \sum_{j>i}\epsilon_{x_i}^2 \epsilon_{x_j}^2\partial_{{x_i}^2}\partial_{{x_j}^2} f /4 \rp$, where terms proportional to $\epsilon$ are not reported since they vanish when integrating. Terms in $\epsilon^2$ become variances when integrating.
We finally substitute the ansatz \cref{eq_lam_exp_app,eq_sb_exp_app} and Taylor expand the right hand side of \cref{eq_self_cons_1} in the small noise limit, which gives constraints on the coefficients $\hat{C}$ and $\hat{D}$. Canceling terms proportional to $\mathrm{CV}_{\alpha}^2$, $\mathrm{CV}_{\varphi}^2$ and $\mathrm{CV}^2_{\mathrm{tree}}[s_b]$ in the right hand side of \cref{eq_self_cons_1} respectively gives
\begin{align}
	\hat{C}_{\alpha} &= - \lp 1-\frac{\ln 2}{2} \rp \\
	\hat{D}_{\varphi} &= - \frac{2-a}{2} \\
	\hat{D}_{s_b} &= \frac{a}{2} \,.
\end{align}
Canceling the terms proportional to  $\mathrm{CV}_{\alpha}^2 \mathrm{CV}_{\varphi}^2$ and $\mathrm{CV}_{\alpha}^2 \mathrm{CV}^2_{\mathrm{tree}}[s_b]$  implies relations between the other coefficients:
\begin{align}
	\label{eq_cstr_cross_terms_1}
	8 \lp (2-a) \hat{D}_{\varphi,\alpha} - 2 \hat{C}_{\alpha,\varphi}  \ln 2 \rp +(2- \ln 2)(2-a)^2&=0 \\
	\label{eq_cstr_cross_terms_2}
	8 \lp (2-a) \hat{D}_{s_b,\alpha} - 2 \hat{C}_{\alpha,s_b}  \ln 2 \rp +(2- \ln 2)(2-a)^2&=0 \,.
\end{align}

To fully determine these last coefficients, we next compute $\langle s_b \rangle_{\mathrm{tree}}$ ($n=1$), using the conservation of volume $\int \di s_b \ s_b \kappa(s_b|s_d') = s_d'/2$: 
\begin{equation}
	\langle s_b \rangle_{\mathrm{tree}} = \int \di s_d' \di s_b' \di \alpha' \ 
	\rho_{\mathrm{tree}}(s_b') \nu(\alpha') \varphi(s_d'-a s_b') s_d' \lp \frac{s_b'}{s_d'} \rp^{\lambda / \alpha'}\,.
\end{equation}
We solve this equation after expanding it for small noise, and canceling the terms proportional to  $\mathrm{CV}_{\alpha}^2 \mathrm{CV}_{\varphi}^2$ and $\mathrm{CV}_{\alpha}^2 \mathrm{CV}^2_{\mathrm{tree}}[s_b]$ respectively implies:
\begin{align}
	\hat{C}_{\alpha, \varphi} &= \frac{(2-a)^2(2+\ln 2)}{16 \ln 2} \\
	\hat{C}_{\alpha, s_b} &= \frac{(2-a)(2(2-a)-(2+a)\ln 2)}{16 \ln 2} \,.
\end{align}
Plugging these values in  \cref{eq_cstr_cross_terms_1,eq_cstr_cross_terms_2} gives:
\begin{align}
	\hat{D}_{\varphi,\alpha}&=\frac{(2-a) \ln 2}{4} \\
	\hat{D}_{s_b,\alpha}&=-\frac{a \ln 2}{4} \,.
\end{align}

Finally, we identify $\mathrm{CV}^2_{\mathrm{tree}}[s_b]$ by computing the second moment ($n=2$) of the birth size distribution:
\begin{align}
	\langle s_b^2 \rangle_{\mathrm{tree}} &= \langle s_b \rangle_{\mathrm{tree}}^2 (1+\mathrm{CV}^2_{\mathrm{tree}}[s_b]) \nn \\
	&= \frac{1+\mathrm{CV}^2_{p}}{2} \int \di s_d' \di s_b' \di \alpha' \ 
	\rho_{\mathrm{tree}}(s_b') \nu(\alpha') \varphi(s_d'-a s_b') s_d'^2 \lp \frac{s_b'}{s_d'} \rp^{\lambda / \alpha'}\,,
\end{align}
where we used $\int \di s_b \ s_b^2 \kappa(s_b|s_d') = s_d'^2 \int \di p \ p^2 \pi(p) = s_d'^2 \langle p \rangle^2 (1+\mathrm{CV}^2_{p})$ with $\langle p \rangle=1/2$.
Expanding this equation for small noise allows us to solve for $\mathrm{CV}^2_{\mathrm{tree}}[s_b]$:
\begin{equation}
	\label{eq_CV_tree_app}
	\mathrm{CV}_{\mathrm{tree}}^2[s_b] \approx  \frac{2-a}{2+a} \mathrm{CV}_{\varphi}^2 + \frac{4}{4-a^2} \mathrm{CV}_{p}^2 \,.
\end{equation}
This expression does not involve first-order term in $\mathrm{CV}_{\alpha}^2$, which was expected since, as shown in the main text, in the absence of noise in size control and volume partitioning all cells have the same birth size along forward lineages, and thus also in the entire tree, so that in this case $\mathrm{CV}_{\mathrm{tree}}^2[s_b]=0$. The expansion also reveals that higher order corrections in $ \mathrm{CV}_{\varphi}^2\mathrm{CV}_{\alpha}^2$ and $ \mathrm{CV}_{p}^2\mathrm{CV}_{\alpha}^2$ do not appear in $\mathrm{CV}_{\mathrm{tree}}^2[s_b]$. Fluctuations in single-cell growth rates might impact $\mathrm{CV}_{\mathrm{tree}}^2[s_b]$ at higher orders, but these are not considered in the expansions \cref{eq_lam_exp_app,eq_sb_exp_app}. Finally, the CV of the birth size distribution is the same in the population and forward lineage statistics (eq. (14) in the main text).

We finally plug \cref{eq_CV_tree_app} into \cref{eq_lam_exp_app,eq_sb_exp_app} and gather the terms in the same way as in eqs. (12) and (15) in the main text:
\begin{align}
	C_{\alpha}&=\hat{C}_{\alpha} \\
	C_{\alpha, \varphi}&=\hat{C}_{\alpha, \varphi}+\frac{2-a}{2+a} \hat{C}_{\alpha, s_b}  \\
	C_{\alpha, p}&= \frac{4}{4-a^2} \hat{C}_{\alpha, s_b} \\
	D_{\varphi}&=\hat{D}_{\varphi}+\frac{2-a}{2+a} \hat{D}_{s_b} \\
	D_{p}&=\frac{4}{4-a^2}\hat{D}_{s_b} \\
	D_{\varphi,\alpha}&=\hat{D}_{\varphi,\alpha}+\frac{2-a}{2+a} \hat{D}_{s_b,\alpha} \\
	D_{p,\alpha}&=\frac{4}{4-a^2} \hat{D}_{s_b,\alpha} \,.
\end{align}

\section{Bi-linear growth of the fission yeast}

The growth pattern of the fission yeast \textit{Schizosaccharomyces pombe} is a long-standing debate, for which exponential \cite{cooper_schizosaccharomyces_2013} and bi-linear \cite{baumgartner_growth_2009,horvath_cell_2016} elongations are popular candidates. To illustrate the effect of growth laws with piecewise patterns on our results, we describe the growth of \textit{S. pombe} with a bi-linear pattern characterized by two portions of linear growth with rates  $\alpha_1$ and $\alpha_2>\alpha_1$, separated by the Rate Change Point (RCP) \cite{baumgartner_growth_2009}. Unlike \textit{Escherichia coli} which elongates by inserting new material along its entire length, \textit{S. pombe} grows by adding material at its tips only \cite{chang_how_2014}. At the beginning of the cycle, \textit{S. pombe} grows only at its old end (tip), and then starts to grow also at its new end, which is called the New End Take Off (NETO) event. It is usually believed, yet still debated \cite{baumgartner_growth_2009}, that RCP and NETO coincide, and that the increase in cell growth rate from $\alpha_1$ to $\alpha_2$ is triggered by the initiation of growth at the new end \cite{horvath_cell_2016}. 

In our framework, bi-linear growth (and similarly for other growth laws with piecewise patterns) enters via the generation time $\tau_d(s_d,s_m,s_b,\alpha_1,\alpha_2)=(s_m-s_b)/\alpha_1+(s_d -s_m)/\alpha_2$, where $s_m$ is the cell size at the RCP, which is constrained by $s_b \leq s_m \leq s_d$. 
Following the idea that RCP and NETO coincide, $\alpha_2$ obeys $\alpha_2=\alpha_1 + \alpha_c + \xi$ with $ \langle \xi \rangle=0$. 
If the old end continues to elongate at the same rate $\alpha_1$ after RCP/NETO as before, then $\alpha_c$ is interpreted as the average elongation rate at the new end, but we do not need to make such assumption here and we simply consider that $\alpha_2$ is the combined elongation rate of the two ends.
Motivated by the recent observation of a size control mechanism for the RCP and the end of the elongation phase \cite{horvath_cell_2016}, indicating a sizer for both $s_m$ and $s_d$ for the Wild Type (WT), and intermediate mechanisms between a sizer and an adder for mutant strains, we define the linear maps $s_m=a_1 s_b + b_1 +\eta_1$ and $s_d=a_2 s_m + b_2 + \eta_2$. The conservation of volume imposes $\langle s_b \rangle_{\mathrm{fw}}  = \langle s_d \rangle_{\mathrm{fw}}  /2= (a_2 b_1 + b_2)/(2-a_1 a_2)$, but does not constrains the average value of $s_m$. 
Finally, the growth parameters $\boldsymbol{\alpha}=(\alpha_1,\alpha_2,s_m)$, together with birth and division sizes, are inherited at division with probability $\mathcal{K}(s_d,s_b,\boldsymbol{\alpha}|s_d',s_b',\boldsymbol{\alpha}') = \varphi_2(s_d-a_2 s_m) \varphi_1(s_m-a_1 s_b) \nu_2(\alpha_2|\alpha_1) \nu_1(\alpha_1) \mathcal{K}_1(s_b|s_d')$. 

Please note that the cell cycle of \textit{S. pombe} is in reality more complex \cite{jia_characterizing_2022}: in addition to the bi-linear growth period (G2 phase), the cycle starts with a rapid but short growth phase corresponding to the formation and shaping of the new end (G1/S phase) and ends with a period of very slow elongation during mitosis and cytokinesis (M phase) \cite{baumgartner_growth_2009}. However, since we wish to illustrate the impact of growth laws with piecewise patterns on population growth, we only consider the bi-linear growth phase for simplicity. Extra phases would be easily modeled by additional parameters $\alpha_i$ and $s_m^i$.

In the absence of noise, the Euler-Lotka equation (eq. (6) in the main text) reduces to $1=2 \exp[-\lambda ((s_m-s_b)/\alpha_1+(2 s_b -s_m)/\alpha_2)]$, which implies $\lambda=\alpha_1 \alpha_2 \ln 2/(s_b(2 \alpha_1 - \alpha_2)+s_m(\alpha_2-\alpha_1))$. Like for single-cell linear growth, the population growth rate explicitly depends on cell birth size $s_b$, and also on the size $s_m$ at the RCP. 
For small noise, the population growth rate takes the form
\begin{equation}
	\label{eq_lambda_bilin}
	\frac{\lambda}{\lambda_0} \approx 1 + \sum_{i=1}^{2} \mathfrak{C}_{\alpha_i} \mathrm{CV}_{\alpha_i}^2 + \sum_{i=1}^{2} \mathfrak{C}_{\varphi_i} \mathrm{CV}_{\varphi_i}^2 + \mathfrak{C}_{p} \mathrm{CV}_{p}^2 \,,
\end{equation}
with $\lambda_0=\langle \alpha_1 \rangle \langle\alpha_2 \rangle \ln 2/(\langle s_b \rangle_{\mathrm{fw}} (2 \langle\alpha_1 \rangle- \langle \alpha_2 \rangle) + \langle s_m \rangle_{\mathrm{fw}} (\langle \alpha_2 \rangle -\langle \alpha_1 \rangle))$, $\mathrm{CV}_{\varphi_i}^2=\langle \eta_i^2 \rangle / b_i^2$ for $i=1,2$, and $\mathrm{CV}_{\alpha_2}^2 = (\langle \alpha_2^2 \rangle - \langle \alpha_2 \rangle^2)/\langle \alpha_2 \rangle^2 = (\langle \alpha_1 \rangle^2 \mathrm{CV}_{\alpha_1}^2 + \langle \xi^2 \rangle)/((\langle \alpha_1 \rangle+\alpha_c)^2)$.

We now define the ratio of growth rates $r_{\alpha}=\langle \alpha_2 \rangle / \langle \alpha_1 \rangle$, called the strength of the RCP \cite{horvath_cell_2016}, and the relative position of the RCP in the cell cycle $r_s=\langle s_m \rangle / \langle s_b \rangle$.
The strength of the RCP $r_{\alpha}$ typically takes values in $1.3 - 1.9$ \cite{horvath_cell_2016}, but for more generality we only assume that $r_{\alpha}>0$. The position of the RCP on the other hand is constrained by $1 \leq r_s \leq 2$, where $r_s=1$ (resp. $r_s=2$) implies simple linear growth with rate $\alpha_2$ (resp. $\alpha_1$) throughout the entire cycle. A value $r_s=1.5$ indicates that the growth rate change occurs halfway through the cycle (in terms of size).

All the coefficients in \cref{eq_lambda_bilin} can be computed by expanding the Euler-Lotka equation (eq. (6) in the main text) as done previously for the other growth laws. To illustrate the impact of the existence of a RCP, we decide to focus on the corrections due to fluctuating single-cell growths rates $\alpha_1$ and $\alpha_2$:
\begin{align}
	\mathfrak{C}_{\alpha_1}&=-\frac{ (r_s -1)(2(2-r_s)(r_{\alpha}-\ln 2)+r_{\alpha}^2(r_s-1)(2-\ln 2))}{2(r_{\alpha}(r_s-1)+(2-r_s))^2} \\
	\mathfrak{C}_{\alpha_2}&=-\frac{ (2-r_s)(2r_{\alpha}(r_s-1)+(2-r_s)(2-\ln 2))}{2(r_{\alpha}(r_s-1)+(2-r_s))^2} \,.
\end{align}
The population growth rate now depends on the strength and the position of the RCP. We recover the corrections computed for linear growth in eq. (22) from the main text when $r_s=1$ (resp. $r_s=2$): $\mathfrak{C}_{\alpha_1}=0$ and $\mathfrak{C}_{\alpha_2}=-(1-\ln(2)/2)$ (resp. $\mathfrak{C}_{\alpha_1}=-(1-\ln(2)/2)$ and $\mathfrak{C}_{\alpha_2}=0$). 

One can now ask whether population growth is more or less sensitive to fluctuations in single-cell growth rate for bi-linear growth than for linear or exponential growths. To make the comparison with linear (or exponential) growth with noise $\mathrm{CV}_{\alpha}$ possible, we consider the case where fluctuations in $\alpha_1$ and $\alpha_2$ are of the same level, i.e. $\mathrm{CV}_{\alpha_1} = \mathrm{CV}_{\alpha_2} \equiv \mathrm{CV}_{\alpha}$, so that the total correction proportional to $\mathrm{CV}_{\alpha}$ is given by $\mathfrak{C}_{\alpha}^{\mathrm{tot}}=\mathfrak{C}_{\alpha_1}+\mathfrak{C}_{\alpha_2}$. This is the case for \textit{S. pombe} WT for which $\mathrm{CV}_{\alpha_1} \approx  0.98 \mathrm{CV}_{\alpha_2}$ \cite{horvath_cell_2016}. The correction $\mathfrak{C}_{\alpha}^{\mathrm{tot}}$ takes values between $-(1-(1+1/r_{\alpha})\ln(2)/4)$, reached for intermediate value $r_s = (2+r_{\alpha})/(1+r_{\alpha})$, and $-(1-\ln(2)/2)$ reached for $r_s=0$ and $r_s=1$. Therefore, fluctuations in single-cell growth rates decreases the population growth further for bi-linear growth than for linear or exponential growths. 
For \textit{S. pombe} WT, reported values $r_{\alpha}=1.61$ and $r_s=1.56$ \cite{horvath_cell_2016} give a correction $\mathfrak{C}_{\alpha}^{\mathrm{tot}} \approx -0.71$.

\section{Population growth rate versus mean birth size in the tree statistics for linearly-growing cells}

\begin{figure*}
	\includegraphics[width=\linewidth]{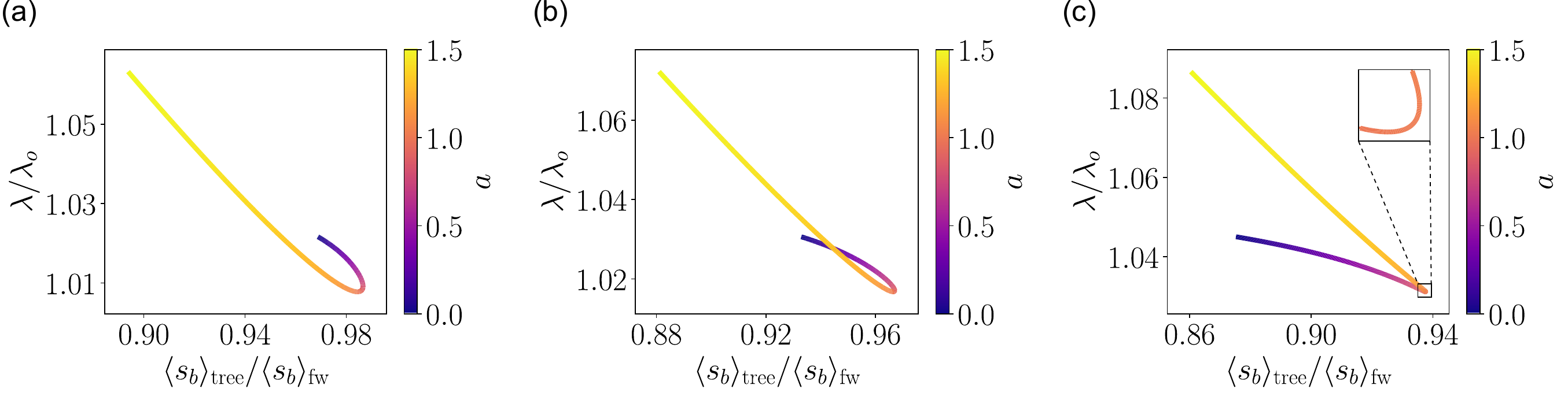}
	\caption{Population growth rate versus mean birth size in the tree statistics for linearly-growing cells. (a-c) $\mathrm{CV}_{\alpha}=0$, $\mathrm{CV}_{p}=0.2$, $b=2-a$ so that $\langle s_b \rangle_{\mathrm{fw}}$ and $\lambda_0$ are kept constant  when varying the mechanism of cell size control $a$. (a) $\mathrm{CV}_{\varphi}=0.15$, (b) $\mathrm{CV}_{\varphi}=0.22$, (c) $\mathrm{CV}_{\varphi}=0.3$.}
	\label{fig_lambda_vs_sb}
\end{figure*}

In the main text, we showed the doubling time of a linearly-growing cell depends on its birth size as: $\tau_2=s_b/\alpha$. Therefore, we expect the population growth rate and the mean birth size in the population to be negatively correlated. 
In this section, we ask whether it is always true that smaller linearly-growing cells lead to faster population growth. To answer this question we plot $\lambda/\lambda_0$ versus $\langle s_b \rangle_{\mathrm{tree}}/\langle s_b \rangle_{\mathrm{fw}}$, respectively given by eqs. (22) and (21) in the main text, when varying the mechanism of cell size control $a$ between $0$ (sizer) and $1.5$ (timer-like) for a fixed value of $\mathrm{CV}_{p}$ and different values of $\mathrm{CV}_{\varphi}$ in \cref{fig_lambda_vs_sb}. We choose $b=2-a$ so that $\langle s_b \rangle_{\mathrm{fw}}$ and $\lambda_0$ are the same for all values of $a$. The graphs are composed of two branches along which the population growth rate and the average birth size evolve in opposite directions. However, $\lambda$ and $\langle s_b \rangle_{\mathrm{tree}}$ can increase or decrease together when going from one branch to the other. Moreover, the two branches are connected by a small region around the adder where $\lambda$ and $\langle s_b \rangle_{\mathrm{tree}}$ evolve in the same direction when varying $a$ continuously. 
We conclude that there exist cases where both the average cell size and the population growth rate evolve in the same direction when modifying the mechanism of cell size control $a$, which goes against our initial intuition.

\section{Additional data analysis}
\label{app_data}

\begin{figure}
	\centering
	\includegraphics[width=0.8\linewidth]{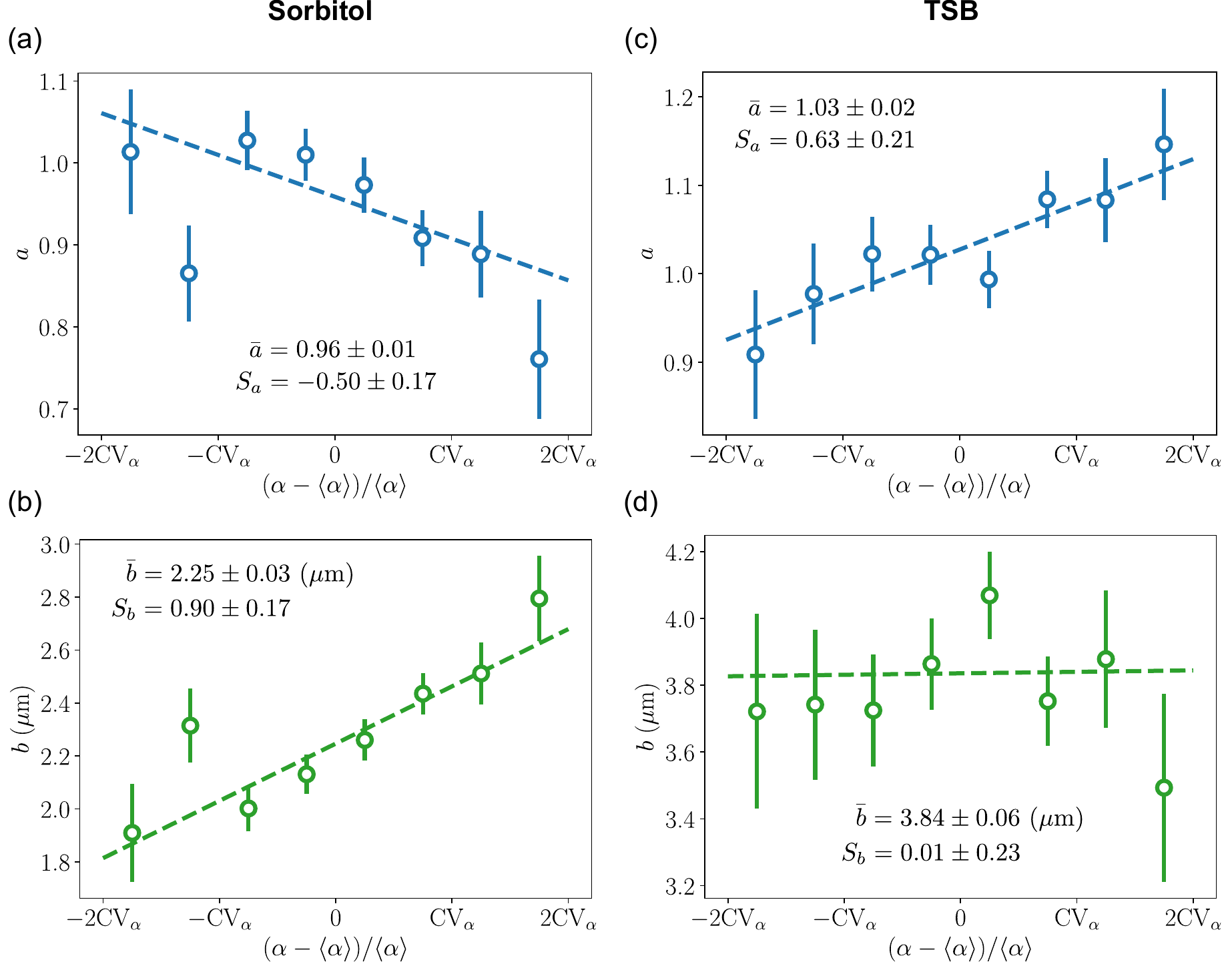}
	\caption{Analysis of mother machine data from Ref. \cite{taheri-araghi_cell-size_2015}. (a,b) \textit{E. coli} in Sorbitol, (c,d) \textit{E. coli} in TSB. (a-d) Single-cell growth rates in an interval of 4 standard deviations centered around the mean, $(\alpha-\langle \alpha \rangle)/\langle \alpha \rangle \in [- 2\mathrm{CV}_{\alpha}, 2\mathrm{CV}_{\alpha}]$, were divided into 8 bins. For each bin the slope $a$ and intercept $b$ were obtained with a linear fit of $s_d$ vs $s_b$. Error bars represent the uncertainties on the parameters $a$ and $b$ of the best linear fit of $s_d$ vs $s_b$. The parameters of the best linear fits of $a$ and $b$ vs $(\alpha-\langle \alpha \rangle)/\langle \alpha \rangle $, shown with dashed lines, are given in the plots.}
	\label{fig:sorb_TSB}
\end{figure}

In this section, we complement the analysis of  mother machine data from Ref. \cite{taheri-araghi_cell-size_2015} for \textit{E. coli} in glucose presented in the main text with two other conditions: Sorbitol and Tryptic Soy Broth (TSB). We show in \cref{fig:sorb_TSB} the dependence of the slope $a$ and intercept $b$ of the noisy linear map $s_d = a s_b + b + \eta$ on single-cell growth rate $\alpha$, for $\alpha$ in an interval of four standard deviations centered around the mean: $(\alpha-\langle \alpha \rangle)/\langle \alpha \rangle \in [- 2\mathrm{CV}_{\alpha}, 2\mathrm{CV}_{\alpha}]$, divided in 8 bins. 

The analysis reveals that in Sorbitol the trend of slope $a$ and intercept $b$ versus single-cell growth rate $\alpha$ are the same as for glucose analyzed in the main text ($S_a<0$ and $S_b>0$). In TSB on the other hand, the slope $a$ increases with $\alpha$ ($S_a>0$) indicating that the faster the cell is growing the closer it gets to a timer. The intercept $b$ is noisy but the best linear fit seems to indicate an independence of the intercept $b$ with the single cell growth rate $\alpha$.

\section{Bounds on the population growth rate}
\label{app_bound}

In this section we explore the validity of the bounds on the population growth rate $\lambda$ derived in \cite{thomas_single-cell_2017} and \cite{genthon_fluctuation_2020} when the target division size $s_d$ depends on the single-cell growth rate $\alpha$ via $\psi(s_d|s_b,\alpha)$. 

For cells growing exponentially at a rate $\alpha = \dot{s}/s$, and when the target division size is independent of the single-cell growth rate, $\psi(s_d|s_b,\alpha)=\psi(s_d|s_b)$, the population growth rate is bounded by \cite{thomas_single-cell_2017}:
\begin{equation}
	\label{eq:PT_bound}
	\langle \alpha^{-1} \rangle^{-1} \leq \lambda \leq \langle \alpha \rangle \,,
\end{equation}
and cannot exceed the mean single-cell growth rate. 

Without any assumption on the details of the model but the existence of a steady-state, we showed that the population growth rate is bounded by (\cite{genthon_fluctuation_2020}, see \cite{genthon_fluctuations_2022} p.35 for details):
\begin{equation}
	\label{eq:my_bound}
	\frac{\ln 2}{\langle \tau_d \rangle_{\mathrm{fw}}} \leq \lambda \,,
\end{equation}
with $\tau_d$ the generation time. 

We start by comparing the two lower bounds for exponentially-growing cells ($\tau_d=\ln(s_d/s_b)/\alpha$) when $\psi(s_d|s_b,\alpha)=\psi(s_d|s_b)$.
When the only source of noise comes from fluctuations in single-cell growth rates, then $\tau_d = \ln 2 / \alpha$ and the two lower bounds in \cref{eq:PT_bound} and \cref{eq:my_bound} are equal. In the presence of other sources of noise however, the two lower bounds differ since $\ln 2 /\langle \tau_d \rangle_{\mathrm{fw}} = \ln 2 \langle \alpha^{-1} \rangle^{-1} \langle \ln(s_d/s_b) \rangle_{\mathrm{fw}}^{-1}$. 
The lower bound in \cref{eq:PT_bound} is saturated in the limit $\mathrm{CV}_{\alpha} \to 0$ regardless of the presence of other noises. 
On the other hand, the statistics of generation times depends on variability in volume partitioning and size control and a small noise expansion of $\langle \tau_d \rangle_{\mathrm{fw}}$ reveals that $\ln 2 /\langle \tau_d \rangle_{\mathrm{fw}} \approx  \langle \alpha^{-1} \rangle^{-1} \Big( 1 - \mathrm{CV}_{p}^2(2-a)/(2 (2+a)\ln 2 - \mathrm{CV}_{\varphi}^2(2-a)^2/(2 (2+a)\ln 2) \Big) $. 
Therefore, the lower bound from \cref{eq:my_bound} is less tight than that from \cref{eq:PT_bound} in the presence of noise in volume partitioning and size control.

We now show that when the target division size $s_d$ depends on the single-cell growth rate $\alpha$ via $\psi(s_d|s_b,\alpha)$, bounds \cref{eq:PT_bound} need not hold while \cref{eq:my_bound} is still valid. Indeed, we see from eq. (30) in the main text that varying the sensitivies $S_a$ and $S_b$ while keeping a fixed noise level $\mathrm{CV}_{\alpha}$ can lead to arbitrary values for the population growth rate. On the other hand, \cref{eq:my_bound} holds since the statistics of generation times depends on $S_a$ and $S_b$. A first-order expansion of the bound indeed reads
\begin{equation}
	\label{eq_low_bound_exp}
	\frac{\ln 2}{\langle \tau_d \rangle_{\mathrm{fw}}} \approx \langle \alpha \rangle \lbk 1 - \Bigg( 1- \frac{1+\ln 2}{2 \ln 2} \lp \Bar{a}S_a + (2-\Bar{a}) S_b \rp + \frac{(\Bar{a}S_a + (2-\Bar{a}) S_b)^2}{2 (2+\Bar{a})\ln 2} \Bigg) \mathrm{CV}_{\alpha}^2 \rbk \,,
\end{equation}
which is smaller that eq. (30) in the main text.

\begin{figure}
	\centering
	\includegraphics[width=0.5\linewidth]{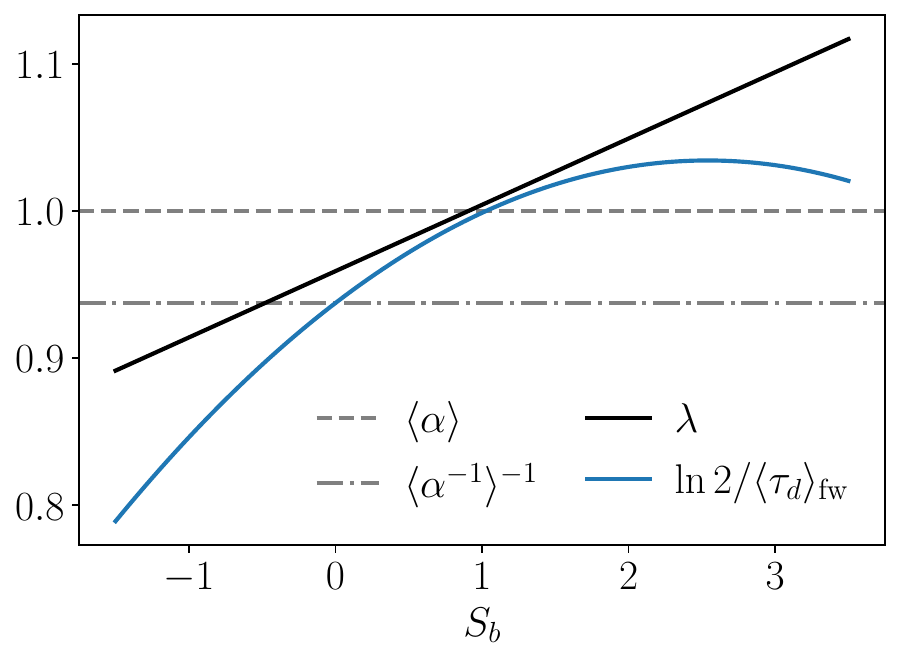}
	\caption{Performance of the bounds on population growth rate $\lambda$ for exponentially-growing cells.  The sensitivity $S_b$ is varied, for an adder $\bar{a}=1$ and $S_a=0$, with fluctuations in single-cell growth rates $\langle \alpha \rangle = 1$ and $\mathrm{CV}_{\alpha}=0.25$. The bounds $\langle \alpha^{-1} \rangle^{-1} \leq \lambda \leq \langle \alpha \rangle$ are not valid, while the lower bound $\ln 2/\langle \tau_d \rangle_{\mathrm{fw}}$, computed with \cref{eq_low_bound_exp} holds for all values of $S_b$.}
	\label{fig:bounds}
\end{figure}

We show in \cref{fig:bounds} the performance of the bounds against $S_b$ for an adder with $\Bar{a}=1$ and $S_a=0$, with variability in single-cell growth rates $\mathrm{CV}_{\alpha}=0.25$ and no noise in size control and volume partitioning. For an adder, the intercept $b$ is the target added volume $\Delta$, so that $S_b=S_{\Delta}$ is the dependence of added volume on single-cell growth rate.

\section{Population growth rate for asymmetrically-dividing cells}

In this section, we show that our formalism can be used to derive the dependence of the population growth rate on the asymmetry of the division obtained in \cite{barber_modeling_2021}. In \cite{barber_modeling_2021}, the authors considered exponentially-growing cells, and assumed no noise in cell size control, a perfect sizer mechanism $a=0$, and an asymmetric but noiseless volume partitioning $\pi(p)=(\delta(p-(1+x)/2) + \delta(p-(1-x)/2))/2$ where $x$ is the division asymmetry and $\delta$ the Dirac delta function. The asymmetry is constrained by $0 \leq x \leq 1$, where $x=0$ gives back symmetric division. 

Since the sizer mechanism is perfect, all cells divide at the same size $s_d=s_d^{\circ}$, so that $\varphi(s_d)=\delta(s_d-s_d^{\circ})$, and therefore the newborn sizes can only take two values: $\rho_{\mathrm{tree}}(s_b)=(\delta(s_b-s_d^{\circ}(1+x)/2) + \delta(s_b-s_d^{\circ}(1-x)/2))/2$. 
The Euler-Lotka equation thus reduces to 
\begin{equation}
	1= \int \di \alpha \ \nu(\alpha) \lp e^{-\frac{\lambda}{\alpha} \ln(\frac{2}{1+x})} + e^{-\frac{\lambda}{\alpha} \ln(\frac{2}{1-x})}\rp \,.
\end{equation}
Expanding this equation for small noise in single-cell growth rates, and using the first-order expansion of the population growth rate in the form of $\lambda / \langle \alpha \rangle \approx 1 +\mathcal{C}_{\alpha} \mathrm{CV}_{\alpha}^2$, gives:
\begin{align}
	1 &= \frac{1+x}{2} + \frac{(1+x)(\ln\lp \frac{1+x}{2}\rp)^2 + 2 (1+x)\ln\lp \frac{1+x}{2}\rp (1+\mathcal{C}_{\alpha})}{4} \mathrm{CV}_{\alpha}^2 \nn \\
	& + \frac{1-x}{2} + \frac{(1-x)(\ln\lp \frac{1-x}{2}\rp)^2 + 2 (1-x)\ln\lp \frac{1-x}{2}\rp (1+\mathcal{C}_{\alpha})}{4} \mathrm{CV}_{\alpha}^2
\end{align}
Finally, canceling the terms proportional to $\mathrm{CV}_{\alpha}^2$ in the right hand side gives the value of $\mathcal{C}_{\alpha}$, and we recover
\begin{equation}
	\frac{\lambda }{\langle \alpha \rangle} \approx 1 - \lp 1 + \frac{1}{2} \frac{(1+x)(\ln\lp \frac{1+x}{2}\rp)^2 + (1-x)(\ln\lp \frac{1-x}{2}\rp)^2 }{(1+x)\ln\lp \frac{1+x}{2} \rp +(1-x) \ln\lp \frac{1-x}{2}\rp } \rp \mathrm{CV}_{\alpha}^2 \,,
\end{equation}
which is the result from \cite{barber_modeling_2021}.